 \documentclass[final,3p,times]{elsarticle2}

\usepackage{amssymb,amsmath}
\usepackage{mathtools}
\usepackage{longtable}
\usepackage{booktabs}
\usepackage[overload]{empheq}
\usepackage{enumitem}
\usepackage{afterpage}
\usepackage{xcolor}
\usepackage{bm}
\usepackage{tikz}

\usepackage{graphicx}
\usepackage{lscape}
\usepackage{caption}
\usepackage{subcaption}
\newcommand*\rfrac[2]{{}^{#1}\!/_{#2}}
\newcommand{\CY}{\ensuremath{c_{y}}~}
\newcommand{\CYc}{\ensuremath{c_{y}}}

\newenvironment{review}{\color{black}}{}
\newcommand{\rev}[1]{\begin{review}#1\end{review}}

\newenvironment{Review}{\color{black}}{}
\newcommand{\Rev}[1]{\begin{Review}#1\end{Review}}

\newenvironment{RReview}{\color{black}}{}
\newcommand{\revrev}[1]{\begin{RReview}#1\end{RReview}}

\journal{Fluids and Structures DOI: 10.1016/j.jfluidstructs.2020.103029}

\begin{document}

\begin{frontmatter}



\title{A nonlinear model of vortex-induced forces on an oscillating cylinder in a fluid flow}


\author[label1,label2]{J. Decuyper}
\ead{jan.decuyper@vub.be}
\author[label1,label2]{T. De Troyer}
\author[label3]{K. Tiels}
\author[label1,label4]{J. Schoukens}
\author[label1,label2]{M. C. Runacres}

\address[label1]{Vrije Universiteit Brussel (VUB), Department of Engineering Technology (INDI), Pleinlaan 2, 1050 Brussels, Belgium.}
\address[label2]{Vrije Universiteit Brussel (VUB), Thermo and Fluid Dynamics (FLOW), Pleinlaan 2, 1050 Brussels, Belgium.}
\address[label3]{Uppsala University (UU), Department of Information Technology, PO Box 337, SE-75105 Uppsala, Sweden.}
\address[label4]{Eindhoven University of Technology (TU/e), Department of Electrical Engineering, Eindhoven, The Netherlands.}

\begin{abstract}
A nonlinear model relating the imposed motion of a circular cylinder, submerged in a fluid flow, to the transverse force coefficient is presented. The nonlinear fluid system, featuring vortex shedding patterns, limit cycle oscillations and synchronisation, is studied both for swept sine and multisine excitation. A nonparametric nonlinear distortion analysis (FAST) is used to distinguish odd from even nonlinear behaviour. The information which is obtained from the nonlinear analysis is explicitly used in constructing a nonlinear model of the polynomial nonlinear state-space (PNLSS) type. The latter results in a reduction of the number of parameters and an increased accuracy compared to the generic modelling approach where typically no such information of the nonlinearity is used. The obtained model is able to accurately simulate time series of the transverse force coefficient over a wide range of the frequency-amplitude plane of imposed cylinder motion.

%
\end{abstract}

\begin{keyword}
Vortex-induced vibrations \sep Forced cylinder oscillations \sep System identification \sep Signal processing \sep Nonlinear black-box modelling \sep Polynomial Nonlinear State-Space model


\end{keyword}

\end{frontmatter}



\section{Introduction}
It is well known that bluff bodies, submerged in a fluid flow, exhibit an unsteady wake for Reynolds numbers exceeding a critical value of $\text{Re}_c\approx 47$. The alternating vortex shedding exerts a periodic forcing on the structure, causing it to vibrate. Such vibrations are generally undesirable since they may accelerate fatigue and even lead to failure. Comprehensive reviews on the mechanism of vortex-induced vibrations are given by \cite{bearman1984,williamson2004,sarpkaya2004}. 

In order to study the flow around oscillating submerged bodies (in this work we will focus on the circular cylindrical shape in a uniform flow) two approaches exist: either the displacement of the body is induced by the flow and the motion is mechanically constrained by springs and dampers, or the displacement is imposed on the body and the resulting flow is studied. Both approaches are respectively known as the freely and the forced vibrating cylinder. Hover et al.\ \cite{hover1998} have, amongst others, shown that the forces acting on freely or forced vibrating cylinders are in close agreement, especially when flow conditions are carefully matched \cite{kumar2016,carberry2004,morse2009}.

Studying the resulting flow from imposed motion offers a number of advantages compared to the free vibration experiment: since the motion is imposed, the studied flow regime can be accurately controlled, moreover, the flow is studied irrespective of any structural parameters, facilitating generalisation of the results. 

It must however be noted that imposing the motion renders mutual interaction between the flow and the displacement impossible. As a result, the intrinsic harmonic features which are present in the fluid force will not be present in the displacement when the motion of the cylinder is imposed. Nevertheless, forced oscillation studies have \rev{led} to an improved understanding of many of the properties of vortex-induced vibrations (VIV) \cite{anagnostopoulos2000_1,carberry2005,khalak1999}.

In many applications where vortex-induced forces are of concern, one requires accurate predictions of the phenomenon. For this complex kind of fluid-structure interaction, an analytical solution can typically not be found. To obtain predictions with high fidelity, either the governing Navier-Stokes equations need to be numerically solved, or an experiment is required. Both approaches are time-consuming and expensive, making them unfit for a large number of intended applications where only a limited amount of resources are available. 

To be of practical use, low-complexity models are needed. Following a semi-empirical approach a number of so called \emph{phenomenological} models were proposed. In \rev{the} early work of Hartlen \& Currie \cite{hartlen1970} a coupled set of ODEs, one of which was nonlinear and of the Van der Pol type, was used. \rev{Splitting the dynamics into a nonlinear model governing the fluid forces and a typically linear counterpart governing the structural behaviour has proven to be a powerful concept. It is within this context that imposed motion experiments become extremely valuable since they allow to model the nonlinear part irrespective of the structural dynamics. Coupling both equations then ensures that only physically relevant regimes are explored.} A number of similar models followed \cite{parkinson1974,facchinetti2004}, with as prime objective to describe and mimic the key features observed from the fluid system. A comprehensive review is given in \cite{gabbai2004}. 

To provide an alternative to CFD or experiments, a higher level of accuracy is required. Promising results have come from the use of system identification techniques. In \cite{runacres2013,cicolani2004} linear, transfer function approaches were applied. Recent work by Zhang et al.\ \cite{zhang2015} provides an excellent illustration of how the classical coupled set of equations, similar to Hartlen \& Currie, can be identified using system identification techniques. The main limitation of their approach follows from the use of a linear model structure \rev{even though} the displacement-to-fluid force relationship is known to be strongly nonlinear \cite{bishop1963}.

A nonlinear modelling approach was presented in \cite{wang2003} where higher harmonic terms, which are evidence of nonlinear behaviour, were recursively introduced. The resulting nonlinear model was able to deal with both cross-flow and inline oscillations. In \cite{decuyper2018} a black-box modelling strategy exploiting nonlinear state-space models was used. Accurate time series simulations of the transverse force coefficient could be reproduced over a wide range of the frequency-amplitude plane of imposed motions. 

The objective of the present work is twofold: to perform a nonlinear analysis of the fluid system by exposing it to a variety of excitation signals (where excitation refers to the imposed motion of the cylinder), and secondly, construct a nonlinear model making explicit use of the insights gained from the nonlinear analysis.

As a quantifier for the fluid system, the transverse force coefficient is used, 
\begin{equation}
\label{e:cy}
c_y(t) = \frac{F_y(t)}{\rfrac{1}{2}\rho U_{\infty}^2},
\end{equation}
with $F_y$ the resultant force in the direction perpendicular to the oncoming flow (a reference frame is shown in Fig.~\ref{f:domain}), $\rho$ the fluid density and $U_{\infty}$ the unperturbed flow speed. \rev{Data are acquired from two-dimensional CFD simulations. Note that given the limitation of a two-dimensional domain, characteristic vortex features such as oblique shedding or spanwise interaction are beyond the scope of the study. }

 \rev{This work is to be interpreted as a first step towards the modelling of freely vibrating cylinders undergoing VIV. The nonlinear model describing the fluid forces constitutes an essential building block of a coupled set of equations \Rev{(similar to \cite{hartlen1970,facchinetti2004,zhang2015})}, governing both the fluid and the structural dynamics. Modelling freely oscillating cylinders is the subject of further research. A proof of concept was provided in \cite{decuyper_phd_2017}.}

The contributions of this work are:
\begin{enumerate}
\item \emph{Analysis of the spectrogram of the fluid system from swept sine excitation.}

This analysis points towards the presence of nonlinear feedback, hence a black-box model structure that includes nonlinear feedback is selected.
\item \emph{Analysis of the type of nonlinearity (odd or even) using the FAST nonparametric distortion analysis.}

A specially designed multisine excitation signal is used to indicate that odd nonlinearities are dominant over even nonlinearities, which is information that is used to limit the degrees of freedom and hence the number of parameters in the black-box model.

\item \emph{Estimation of a black-box polynomial nonlinear state-space (PNLSS) model that includes the information of both analyses.}

This results in a nonlinear model with less parameters and about half the error of the model obtained using the generic approach, presented in \cite{decuyper2018}, which does not use the information obtained from the FAST approach.
\end{enumerate}



The layout of the paper is as follows: we start by introducing the CFD method and its validation which are \rev{used to} acquire the necessary data of the fluid system. Next, the key property of nonlinear systems to generate output power at harmonic frequency lines of the excitation frequency is illustrated on the basis of a swept sine excitation experiment. Then the FAST approach, enabling characterisation of the nonlinear behaviour, is applied to the fluid system. This is followed by the identification of a nonlinear model, tailored using the insight of the FAST test, and a discussion of the results. We conclude by evaluating the benefit achieved from the adopted modelling approach.

\section{The CFD method}

All simulations are performed in the open source CFD environment OpenFOAM \cite{openFoam}. OpenFOAM implements a finite-volume formulation of the discretised Navier-Stokes equations. Version 2.3.1 was used in all cases. 
%

The considered regime is characterised by a constant Reynolds number of $\text{Re} =100$, ensuring laminar, predominantly 2-dimensional vortex shedding. As Strouhal frequency, $f_{\text{St}}=3$ Hz was selected. Fixing both the Reynolds number and $f_{\text{St}}$, effectively defines the unperturbed uniform flow speed \cite{strouhal1878}, provided that the diameter is set ($D=0.10$ m in this case), giving
\begin{equation}
\label{e:Strouhal}
U_{\infty}=\frac{f_{\text{St}}D}{\text{St}}.
\end{equation}
With the Strouhal number \rev{(which is a mild function of Re) $\text{St}=0.167$ for $\text{Re}=100$}, one obtains $||\vec{U}||=U_{\infty}=1.7964$ m/s. To satisfy $\text{Re}=100$ a kinematic viscosity $\nu=1.7964\times 10^{-3}~\text{m}^2/\text{s}$ is selected.



The conservation equations, integrated over the control volumes, are solved using a second order upwind Gauss linear interpolation for the convection term while for the diffusion term a central difference second order Gauss linear interpolation scheme is applied. Time is discretised in a first order Eulerian (implicit) manner. 

The coupled set of continuity and momentum equations are solved using the \emph{pimpleDyMFoam} solver, i.e.\ a transient solver which allows for relatively large time steps thanks to the hybrid PISO-SIMPLE (PIMPLE) algorithm \cite{openFoam}. To ensure (numerical) convergence the Courant number was constrained to $\text{Co}_{\text{max}} \le 0.5$.

\begin{figure}
\begin{center}
\setlength{\unitlength}{0.2cm}
\begin{picture}(42.5,32.5)
        \put(2.5,2.5){\line(1,0){40}}
        \put(42.5,2.5){\line(0,1){30}}
        \put(42.5,32.5){\line(-1,0){40}}
        \put(2.5,32.5){\line(0,-1){30}}
        \put(12.5,17.5){\circle{1}}
        
        \put(22.5,0.5){\vector(-1,0){20}}
        \put(22.5,0.5){\vector(1,0){20}}
        \put(-1.5,17.5){\vector(0,1){15}}
        \put(-1.5,17.5){\vector(0,-1){15}}       
        \put(2.5,0){\line(0,1){1}}
        \put(42.5,0){\line(0,1){1}}       
        \put(-2,2.5){\line(1,0){1}}
         \put(-2,32.5){\line(1,0){1}}
         \put(7.5,16){\vector(-1,0){5}}
         \put(7.5,16){\vector(1,0){5}}
         \put(12.5,15.5){\line(0,1){1}}
         \put(27.5,16){\vector(-1,0){15}}
         \put(27.5,16){\vector(1,0){15}}

        \put(-0.5,32.5){\vector(1,0){2.5}}
        \put(-0.5,30){\vector(1,0){2.5}}
        \put(-0.5,27.5){\vector(1,0){2.5}}
        \put(-0.5,25){\vector(1,0){2.5}}
        \put(-0.5,22.5){\vector(1,0){2.5}}
        \put(-0.5,20){\vector(1,0){2.5}}
        \put(-0.5,17.5){\vector(1,0){2.5}}
        \put(-0.5,15){\vector(1,0){2.5}}
        \put(-0.5,12.5){\vector(1,0){2.5}}
        \put(-0.5,10){\vector(1,0){2.5}}
        \put(-0.5,7.5){\vector(1,0){2.5}}
        \put(-0.5,5){\vector(1,0){2.5}}
        \put(-0.5,2.5){\vector(1,0){2.5}}
        
        \put(43,32.5){\vector(1,0){2.5}}
        \put(43,30){\vector(1,0){2.5}}
        \put(43,27.5){\vector(1,0){2.5}}
        \put(43,25){\vector(1,0){2.5}}
        \put(43,22.5){\vector(1,0){2.5}}
        \put(43,20){\vector(1,0){2.5}}
        \put(43,17.5){\vector(1,0){2.5}}
        \put(43,15){\vector(1,0){2.5}}
        \put(43,12.5){\vector(1,0){2.5}}
        \put(43,10){\vector(1,0){2.5}}
        \put(43,7.5){\vector(1,0){2.5}}
        \put(43,5){\vector(1,0){2.5}}
        \put(43,2.5){\vector(1,0){2.5}}
        
        \put(12.5,17.5){\vector(1,0){5}}
        \put(12.5,17.5){\vector(0,1){5}}
        
        \put(17.5,16.5){\makebox(1,1){$x$}}
        \put(11.25,22){\makebox(1,1){$y$}}
        
        \put(22.5,-1){\makebox(1,1){$L$}}
        \put(-3,17.5){\makebox(1,1){$H$}}
        
         \put(7.5,14){\makebox(1,1){$X_i$}}
         \put(27.5,14){\makebox(1,1){$X_o$}}

\end{picture}
\end{center}
\caption{Topology of the computational domain (scale drawing).}
\label{f:domain}
\end{figure}
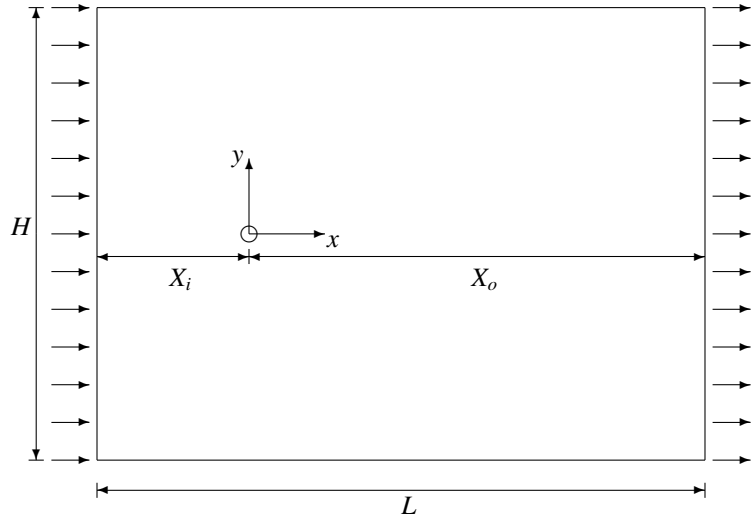

The topology of the 2-dimensional computational domain is shown in Fig.\ \ref{f:domain}. The cylinder is positioned $10D$ from the inlet ($X_i$), $25D$ from the outlet ($X_o$) and is centred in the $y$-direction ($H=30D$). The mesh was constructed using Hexpress (unstructured full-hexahedral meshing) of Numeca.

In the neighbourhood of the cylinder a radial refinement region is applied with a radius of $5D$. The first grid point (cell centre) is located at a relative height from the cylinder surface of $\delta = 0.005D$. This is in agreement with \cite{qu2013}. 

At the inlet, a free stream boundary condition is applied: if $\vec{U}=(u,v,w)$ then $u=U_{\infty}$ and $v=w=0$ (Dirichlet condition) combined with a zero pressure gradient, normal to the inlet, $\frac{\partial p}{\partial x}=0$. At the cylinder surface, the no-slip condition is applied: $u=w=0$ and $v=\dot{y}$, with $y$ the imposed cylinder motion and $\vec{\nabla}_n{p}=\vec{0}$ (subscript $n$ indicating the normal direction). To the lateral boundaries, the slip condition is applied: $v=0$ and $\frac{\partial{u}}{\partial{y}}=0$. Since it is a 2-dimensional domain, the momentum equation is not defined in the third dimension and hence does not need boundary conditions (front and back). 

For the outflow boundary, equivalent to \cite{placzek2009}, a zero normal velocity gradient ($\vec{\nabla}_n{\vec{U}}=\vec{0}$) is imposed (Neumann condition) and the pressure is set to the static pressure, $p=0$ (similar to \cite{pan2007}). 

%
%

\rev{Validation of the CFD model is done on the basis of quantities of interest and is reported in \ref{s21:CFD_validation}.}

\subsection{Vortex patterns at Reynolds number 100}
\begin{figure}
\begin{center}
\includegraphics[scale=0.5]{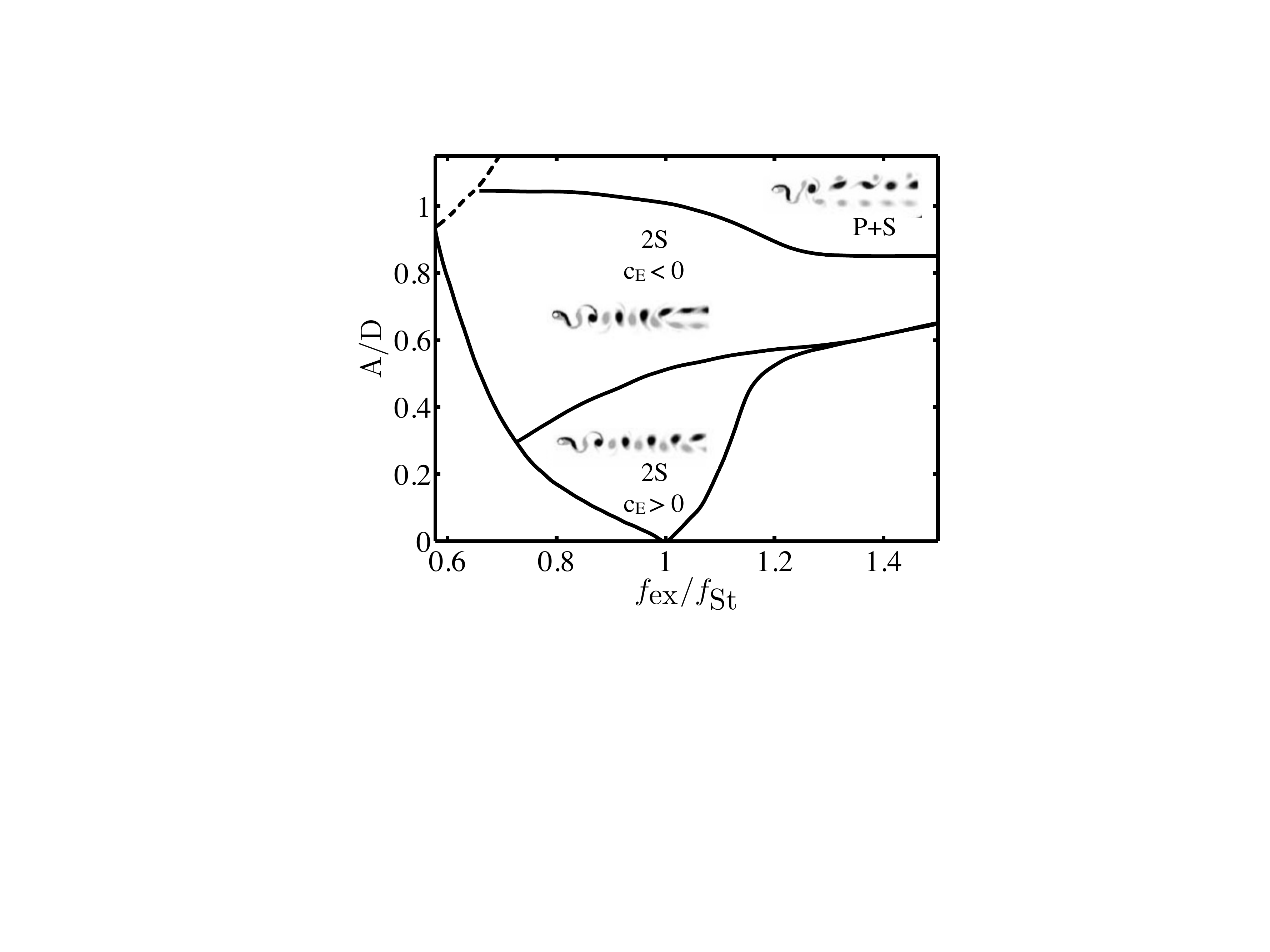}
\caption{Vortex modes in the wake of imposed oscillating cylinders in a cross flow at $\text{Re}=100$. The energy transfer coefficient, denoted with $c_\text{E}$, is positive if energy is transferred from the fluid to the structure and negative otherwise. This figure is a reproduction of Fig.\ 4a from \cite{leontini2006}.}
\label{f:wake_map_Re100}
\end{center}
\end{figure}

\begin{figure}
\begin{center}
\begin{subfigure}[b]{0.48\textwidth}
\includegraphics[width=\textwidth]{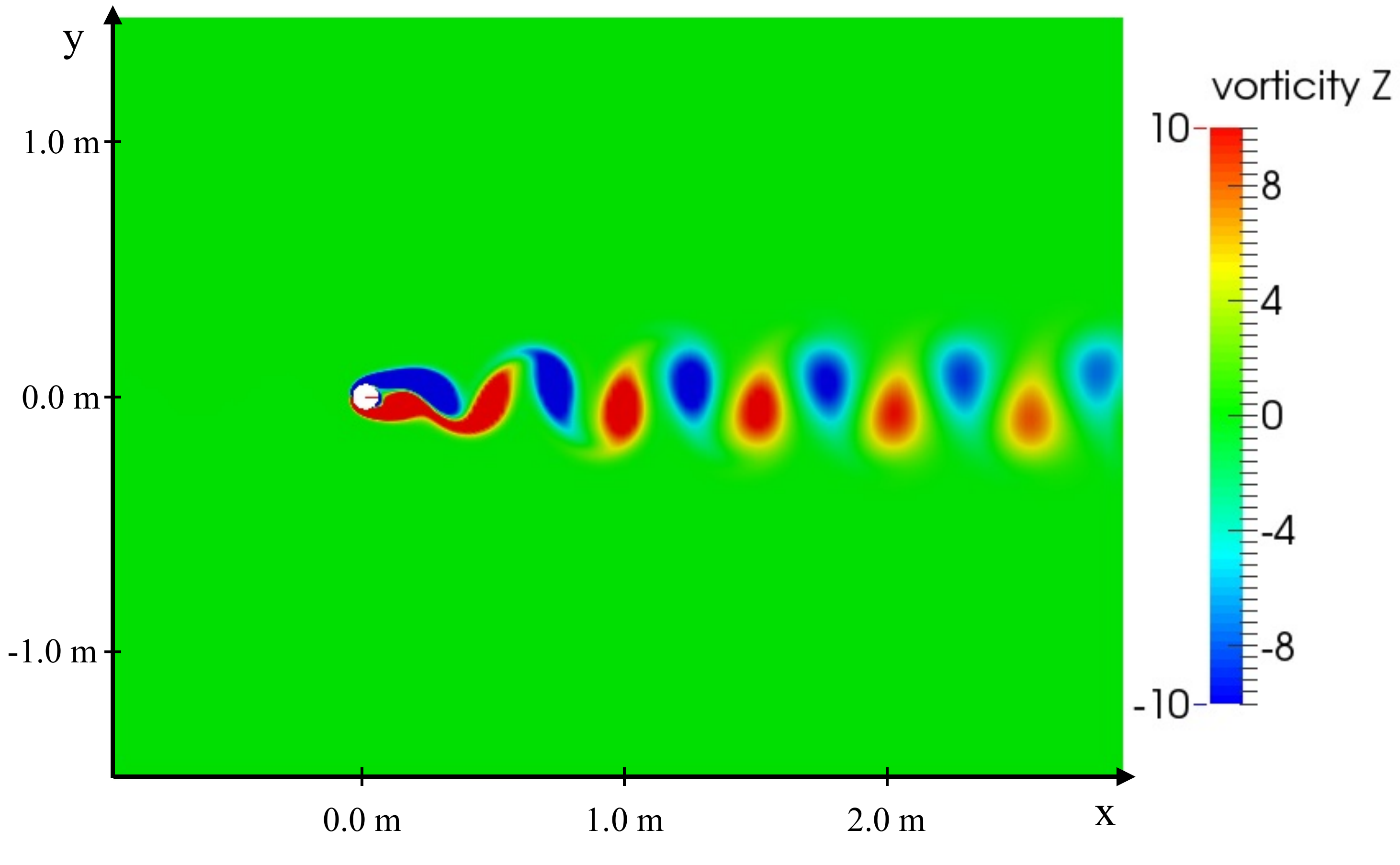}
\caption{}
\end{subfigure}
\begin{subfigure}[b]{0.48\textwidth}
\includegraphics[width=\textwidth]{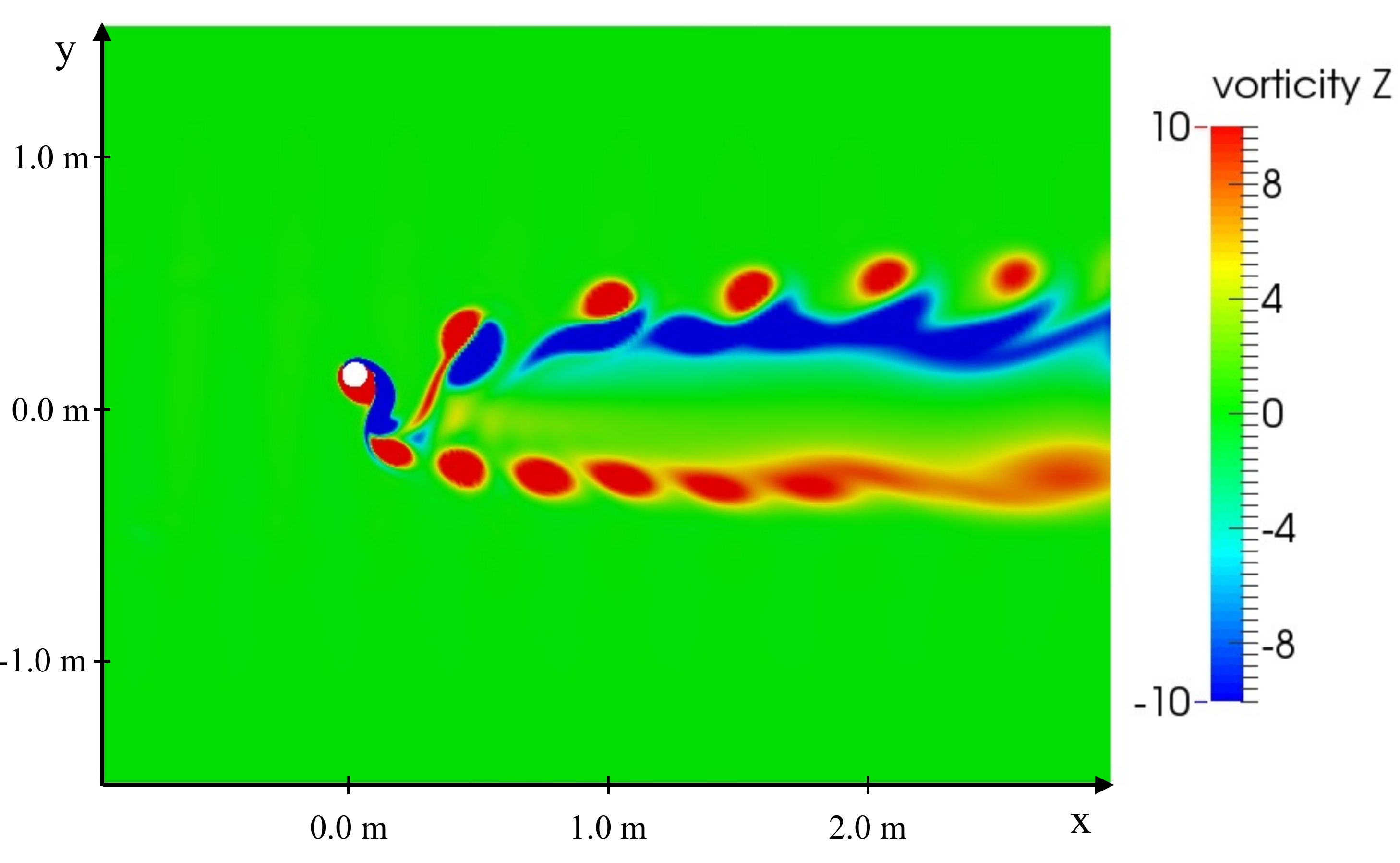}
\caption{}
\end{subfigure}
\caption{Visualisation of CFD simulations at $\text{Re}=100$. (a) Vorticity of the flow field about a stationary cylinder (2S vortex pattern). (b) Vorticity of the flow field about a cylinder undergoing an imposed motion with $f_{\text{ex}}/f_{\text{St}}=1.5$ and $A/D=1.25$. Clockwise vorticity is negative.}
\label{f:CFD_vort}
\end{center}
\end{figure}

In Fig.\ \ref{f:wake_map_Re100} a map of wake patterns is drawn for $\text{Re}=100$. At this Reynolds number, two types of vortex formation are observed \cite{leontini2006}: the 2S mode which displays the shedding of two opposite spinning vortices per cycle and the P+S mode where both a pair and a single vortex are shed per cycle. By plotting the vorticity of the flow field from generated simulations, both modes are identified (Fig.~\ref{f:CFD_vort}). The frequency of imposed motion is denoted by $f_{\text{ex}}$ and the amplitude $A$ is given relative to $D$.

%
%
%

\section{Nonlinear properties of the fluid system}

In this section we study the behaviour of \CY while applying broadband excitation signals as the motion imposed on the cylinder. The use of broadband signals has proven to be beneficial from a modelling perspective. Especially because of the rich and condensed information content, swept sine and multisine signals (see Section \ref{ss:FAST}) are in many cases the preferred excitation signals when working with \rev{data-driven} modelling approaches \cite{pintelon2001}, \Rev{e.g.\ Zhang et al.\ trained their fluid model on the basis of a chirp signal \cite{zhang2015}.}

The objective is to study data of \CYc, and examine them especially for the presence of nonlinear effects. From a system identification perspective one would like to acquire knowledge of the characteristic behaviour of the nonlinearity in order to make an appropriate choice of model structure. Typical questions are: does the output depend on past values of the output itself, or in other words, is the output fed back through the system? And can the type of nonlinear behaviour be characterised as a dominantly odd or even function? Holding the answer to both these questions enables to discriminate between the vast amount of nonlinear model structures that may be applied.

\subsection{Imposing swept sine oscillations}
\label{ss:SWS}
The imposed motion applied here is a swept sine signal. The final time step of the flow field obtained about a stationary cylinder is taken as initial condition. The cylinder is then forced to follow a certain swept-sine trajectory in the $y$-dimension.

\begin{figure}
\begin{center}
\includegraphics[width=\textwidth]{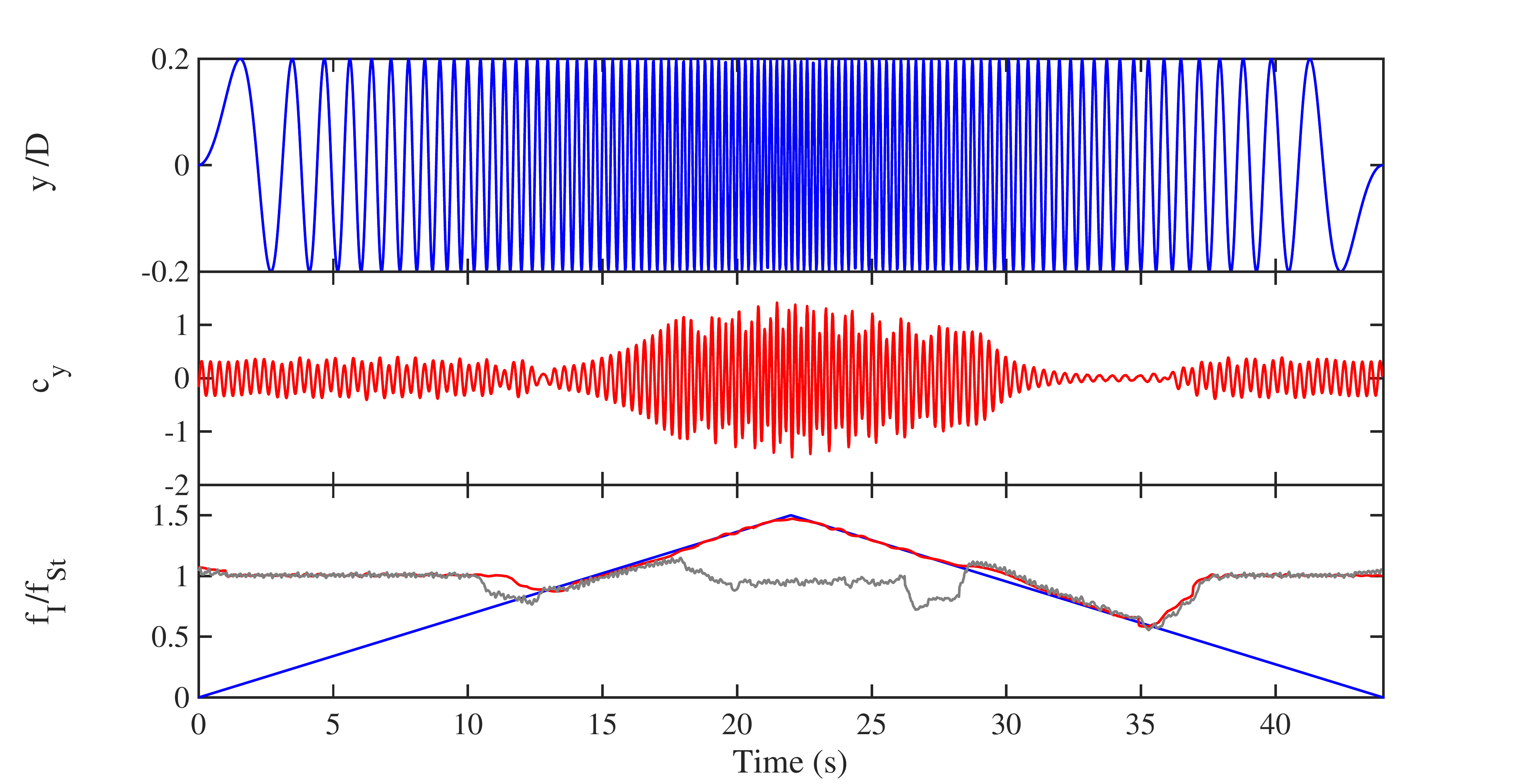}
\caption{Top: time series of the sweep up and sweep down swept sine imposed cylinder motion (in meters). The frequency band covered ranges from $f_{\text{ex}}/f_{\text{St}}=0 - 1.5$ at $A/D=0.20$ and a sweep period (single sweep direction) of $T_0=22$ s. Middle: Time series of the corresponding \CYc. Bottom: instantaneous frequency of the imposed motion (blue), \CY (red) and the vorticity measured in a point $3D$ downstream of the \color{black}{cylinder in grey}.}
\label{f:SWS_time_fI} 
\end{center}
\end{figure}

In Fig.~\ref{f:SWS_time_fI} the time series and the instantaneous frequency of the imposed motion, the measured \CY and the measured vorticity in a point $3D$ downstream of the cylinder are shown. The instantaneous frequency $f_\text{I}$ is calculated as
\begin{equation}
f_{\text{I}}(t)=\frac{1}{2\pi}\frac{d\phi (t)}{dt},
\end{equation}
where $\phi$ denotes the instantaneous phase which is found through the calculation of the Hilbert transform \cite{marple1999}. Untreated, $f_{\text{I}}$ is very sensitive to irregularities in the signal. Moreover an instantaneous frequency is only meaningfully defined for single harmonic signals. The presence of multiple frequency components in the signal distort $f_{\text{I}}$. A moving average technique (central differencing) is therefore applied to $f_{\text{I}}$.

From studying $f_{\text{I}}$ in Fig.~\ref{f:SWS_time_fI}, two periods are identified where the frequency of the $z$-component of the vorticity ($\omega_z$) synchronises to the excitation frequency. The synchronisation is observed both during the sweep up ($13 \le t \le 17$) and sweep down ($28 \le t \le 35$). The property of the unsteady wake to synchronise in frequency to the excitation frequency for certain ranges in oscillation amplitude and frequency is referred to as entrainment and the corresponding range of parameter space as the lock-in region. Comparing \CY during sweep up and sweep down reveals a clear dependency on the sweep direction. This non-uniqueness of \CY for identical frequency and amplitude of input signal yet opposite sweep direction implies that the system behaves nonlinearly in this regime.

\begin{figure}
\begin{center}
\begin{subfigure}[b]{\textwidth}
\begin{center}
\includegraphics[width=0.8\textwidth]{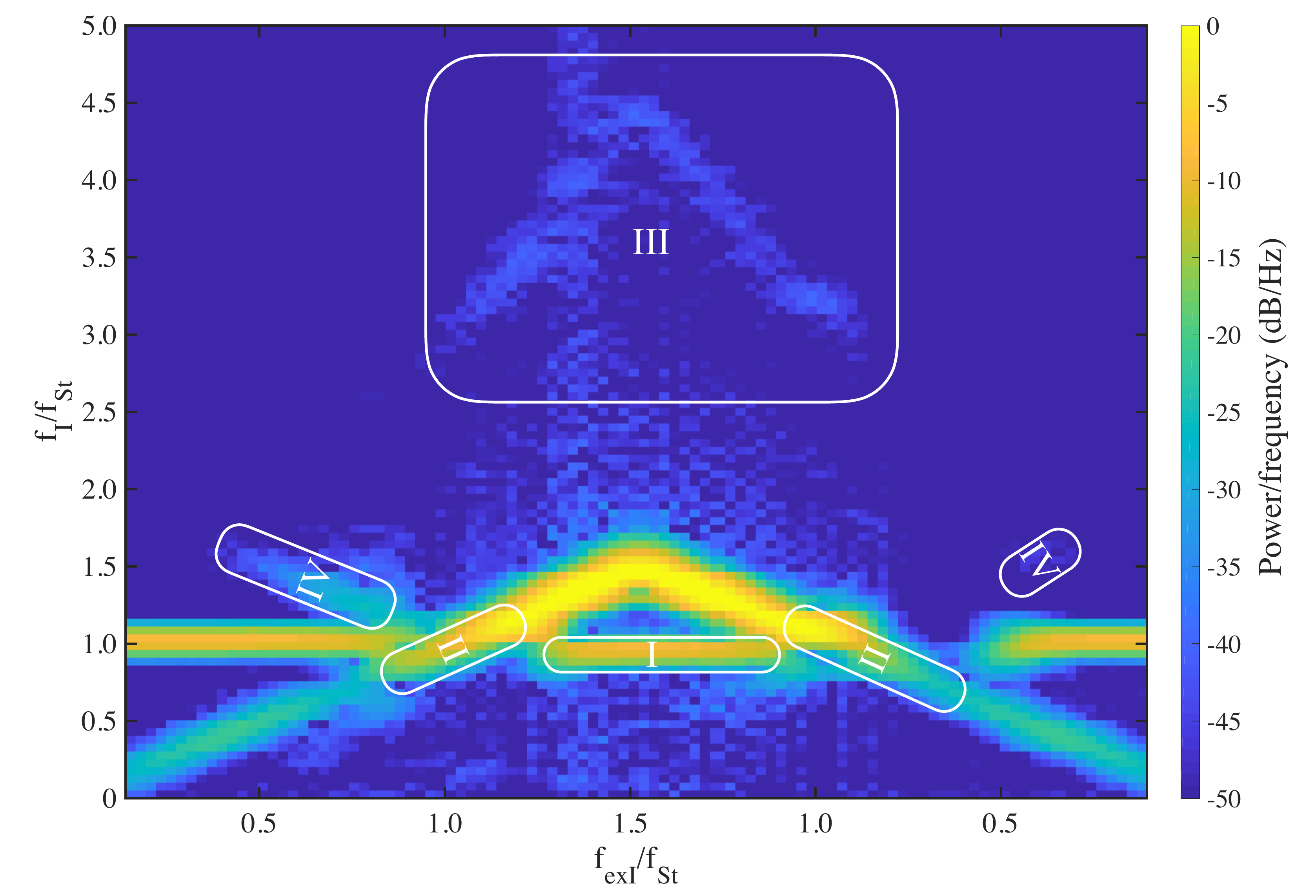} 
\caption{}
\label{f:cy_0p20_spectro}
\end{center}
\end{subfigure}
\begin{subfigure}[b]{\textwidth}
\begin{center}
\includegraphics[width=0.8\textwidth]{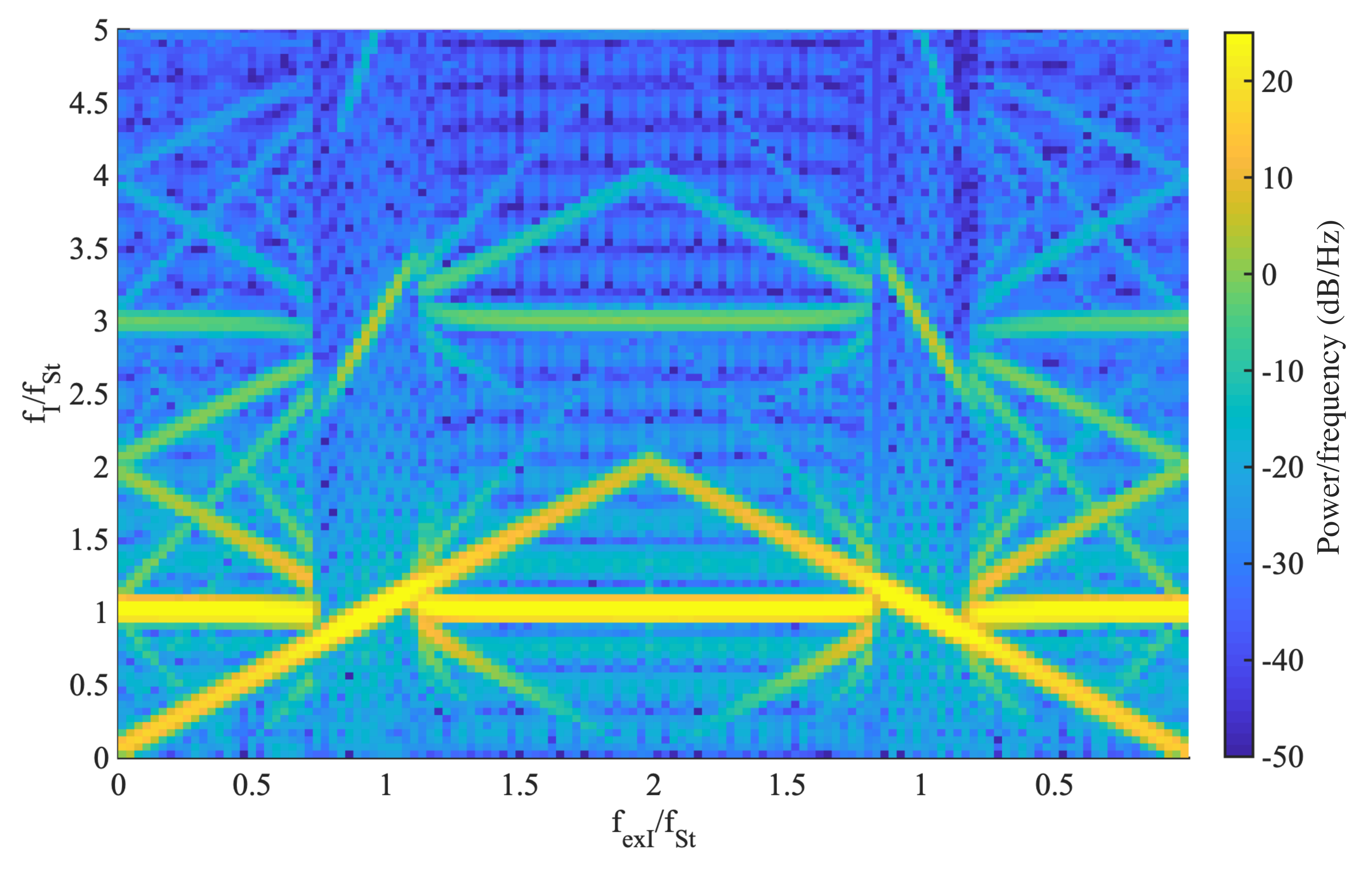}
\caption{}
\label{f:spectro_VdP}
\end{center}
\end{subfigure}
\caption{(a) Spectrogram of \CY for an imposed sine-sweep excitation of $A/D=0.20$ and a sweep period of $T_0=22$ s (similar to Fig.\ \ref{f:SWS_time_fI}). From left to right sweep up is followed by sweep down. \color{black}{(b) Spectrogram of $c_y(t)$ modelled by Eq.~\eqref{e:VdP} when applying a sine sweep (sweeping up and down from left to right) as the right hand side. The instantaneous excitation frequency of the imposed motion is reported on the $x$-axis. Equivalently the simulation time could have been used.} \color{black} Similar observations could be made for relative amplitude levels other than $A/D=0.2$.}
\end{center}
\end{figure}

Given the pronounced difference between the $f_{\text{I}}$ of \CY and $\omega_z$, lock-in requires a more rigorous definition in order to be uniquely defined. In \cite{kumar2016} the wake is considered locked whenever the following two conditions are met:
\begin{enumerate}
\item the dominant frequency in the power spectrum of \CY matches the cylinder oscillation frequency ($f_{\text{ex}}$)
\item if any other power is present in the signal of \CYc, it can only be present at integer multiples of $f_{\text{ex}}$ (higher harmonics).
\end{enumerate}
From the above it follows that only studying the $f_{\text{I}}$ of \CY is insufficient to be able to judge whether or not the wake is locked. Monitoring whether both conditions are met during the sweep experiment can be achieved by studying a spectrogram visualisation. Fig.~\ref{f:cy_0p20_spectro} shows the spectrogram that corresponds to the data of \CY shown in Fig.~\ref{f:SWS_time_fI}. \rev{Similar to Fig.\ \ref{f:SWS_time_fI}, sweep up is followed by sweep down, reading from left to right. In this case the $x$-axis denotes the instantaneous excitation frequency rather than the evolved time of the sweep experiment.}
Four interesting features are indicated by roman numbers. 
\begin{enumerate}[label=\Roman*.]
\item This is the high excitation frequency range ($1.2<f_{\text{exI}}/f_{\text{St}}<1.5$). Within this range it is clear that the second condition, required for lock-in, is not met since there is a clear contribution of power present at $f_{\text{St}}$, which is then a non-integer multiple of the excitation frequency. This is moreover confirmed by the deviation of the $f_{\text{I}}$ of $\omega_z$ from $f_{\text{ex}}$ over the same frequency range of excitation, which can be seen from Fig.~\ref{f:SWS_time_fI}.

\item In these regions the wake, and hence \CYc, synchronise in frequency to the excitation frequency (lock-in regions). \rev{The spectrogram does not show a power concentration other than the one at the excitation frequency. Therefore both conditions for lock-in are met.}

\item This feature reveals the presence of a third harmonic in \CYc. The third harmonic is especially pronounced at high excitation frequencies. Harmonics are a clear evidence of nonlinearity.

\item In this range, \CY contains a contribution which is present at the excitation frequency, mirrored about the Strouhal frequency\footnote{The presence of feature IV is less pronounced during sweep down than during sweep up. Study of the spectrogram at other amplitude levels and sweep rates however confirmed the existence independent of the sweep direction.}. The origin of this specific feature is attributed to the combination of nonlinearity and internal feedback of the output within the system. The following is clarified in the next section by studying the frequency spectrum of \CY along horizontal cuts of the spectrogram. A cut along the swept sine corresponds \rev{to an experiment where the cylinder describes a simple harmonic motion.}

\end{enumerate}

\rev{Fig.~\ref{f:spectro_VdP} depicts the spectrogram of $c_y(t)$ following a swept sine excitation, modelled by a Van der Pol equation of the following form:
\begin{equation}
\ddot{c}_y(t) + \varepsilon \Omega_{\text{St}} \left(c^2_y(t)-1\right) \dot{c}_y(t) + \Omega_{\text{St}}^2c_y(t) = y(t),
\label{e:VdP}
\end{equation}
with parameters chosen as $\varepsilon = 0.3$ and $\Omega_{\text{St}} = 2\pi f_{\text{St}}$ with $f_{\text{St}}=5$ Hz. A sine, sweeping linearly in frequency from $0$ to $2f_{\text{St}}$ and back down to $0$ with an amplitude equal to 500 was applied as forcing term, $y(t)$. The sweep rate was set to $0.012f_{\text{St}}$ per second.

Fig.~\ref{f:spectro_VdP} illustrates clearly that the Van der Pol equation is able to reproduce all features which were observed in the spectrogram of the CFD data (Fig.~\ref{f:cy_0p20_spectro}). It is therefore crucial to determine the basis for the succes of the Van der Pol equation, so that a model structure can be selected which reflects this insight. Section \ref{ss:nonlinear_feedback} attempts to shed light on the matter.
}
\subsection{Nonlinear feedback within the fluid system}
\label{ss:nonlinear_feedback}

\rev{Studying the origin of feature IV of Section \ref{ss:SWS} will allow us to infer the key characteristics that a model structure should posses in order to be able to model the data. A closer look is provided by means of Figures \ref{f:ss_spectro_cut_a} and \ref{f:ss_spectro_cut_b} where the spectra of $c_y(t)$ are shown for a number of single-harmonic imposed motions. One observes that feature IV corresponds to a frequency component at $2f_{\text{St}}-f_{\text{ex}}$. We will show that such a contribution corresponds to a third harmonic of a signal bearing power at both the excitation frequency and the Strouhal frequency.

\thispagestyle{empty}
\begin{figure}
\begin{center}
\begin{subfigure}[b]{\textwidth}
\begin{center}
\includegraphics[width=0.92\textwidth]{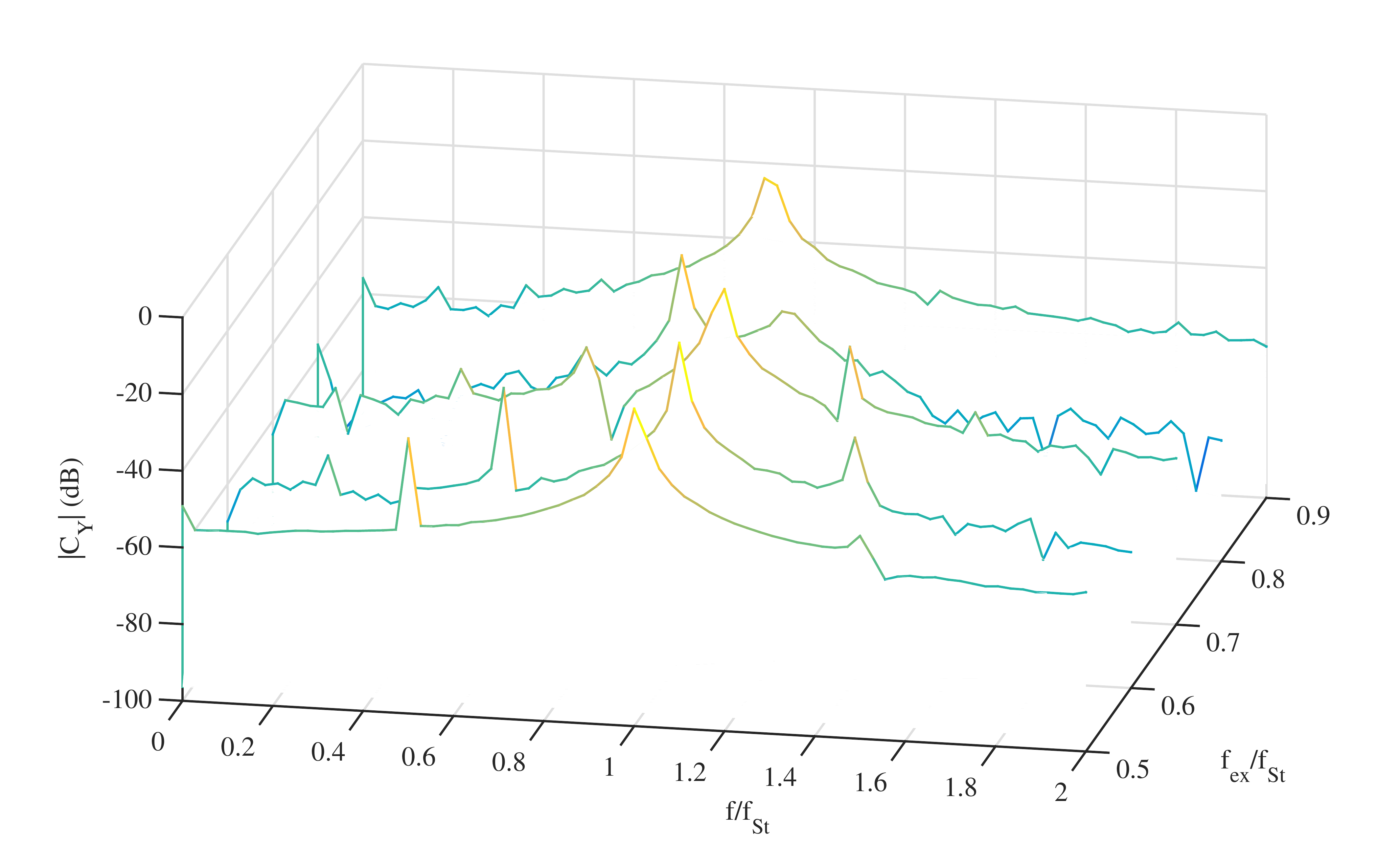}
\caption{}
\label{f:ss_spectro_cut_a}
\end{center}
\end{subfigure}
\begin{subfigure}[b]{\textwidth}
\begin{center}
\includegraphics[width=0.92\textwidth]{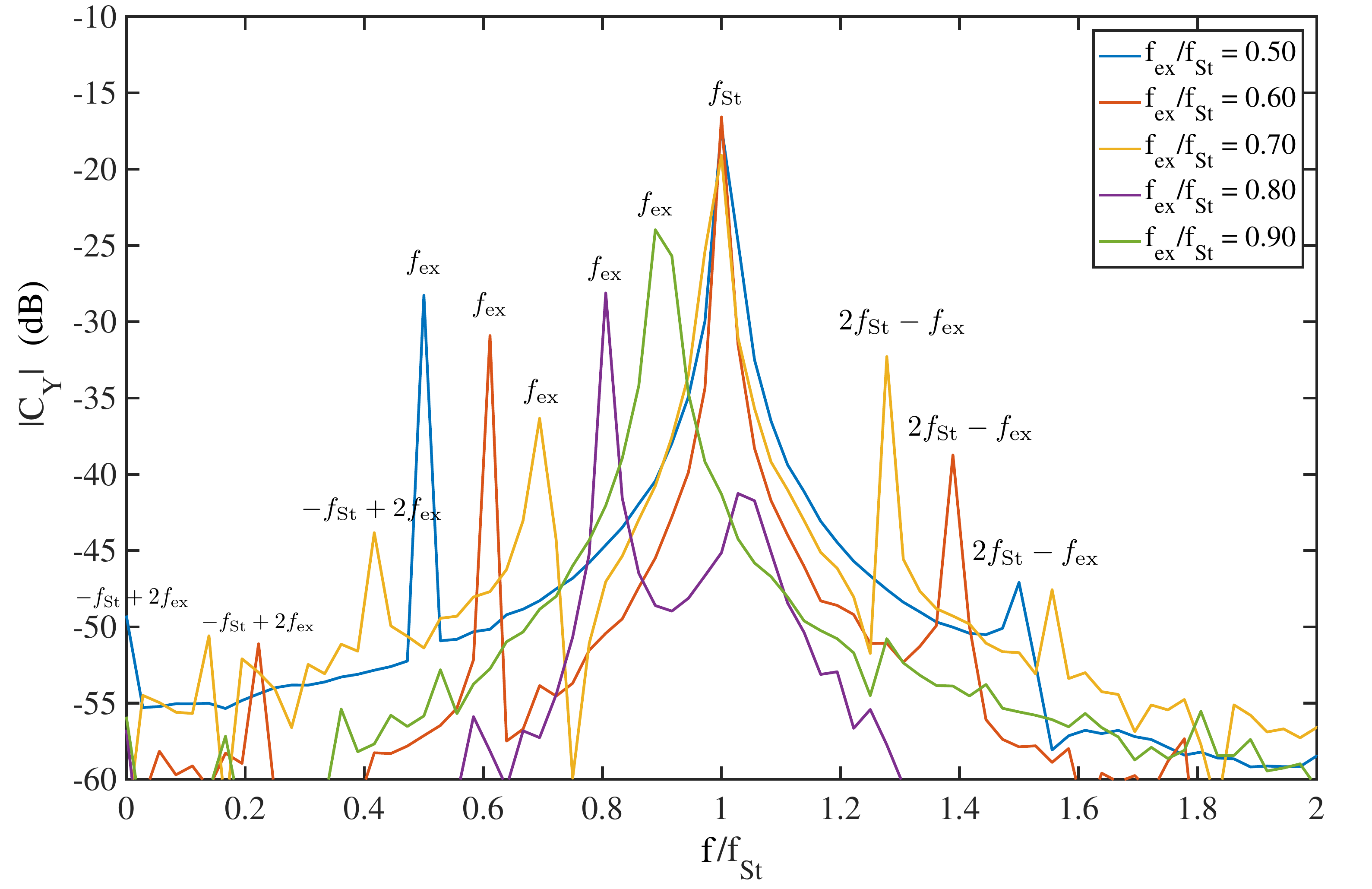}
\caption{}
\label{f:ss_spectro_cut_b}
\end{center}
\end{subfigure}
\caption{DFT spectra of \CY obtained from single harmonic sine excitations at discrete excitation frequencies throughout the frequency range which is indicated by the roman number IV in Fig.~\ref{f:cy_0p20_spectro}. In this region, outside lock-in, output power is observed at the excitation frequency, the Strouhal frequency and at combined frequency lines. This suggests that the system contains a nonlinear feedback loop. (a) waterfall representation. (b) overlay visualisation.}
\label{f:ss_spectro_cut}
\end{center}
\end{figure}

Since we know that the Van der Pol equation is able to reproduce power at $2f_{\text{St}}-f_{\text{ex}}$ (see Fig.~\ref{f:spectro_VdP}) it suffices to examine the structure of Eq.~\eqref{e:VdP}. Expanding the damping term of \eqref{e:VdP} and moving the nonlinear term to the right hand side results in:
\begin{equation}
\label{e:VdP2}
\ddot{c}_y(t) -\varepsilon \Omega_{\text{St}} \dot{c}_y(t) + \Omega_{\text{St}}^2 c_y(t) = y(t) - \varepsilon \Omega_{\text{St}} c^2_y(t)\dot{c}_y(t).
\end{equation}
\begin{figure}
\begin{center}
\setlength{\unitlength}{0.62cm}
\begin{tikzpicture}
        \thinlines
       \node at (2.46,3.5) {\footnotesize{unstable linear system}};
        \put(2,3){\framebox(4,2){}}
        \node at (1.75,2.73) {\small{$h(t)$}};
        
       \begin{scope}[shift={(2.2,0)}]
 \draw [black, domain=0:2] plot [smooth,scale=0.39] (\x, {0.1*exp(1.5*\x)*sin(8*2*pi*\x r)+6.35});
  \end{scope}
  \draw [->] (2.19,2.46) -- (3.2,2.46);
  \draw [-,red] (2.52,2.25) -- (2.62,2.25);
  \draw [-,red] (2.52,2.243) -- (2.52,2.28);
   \draw [-,red] (2.62,2.243) -- (2.62,2.28);
  \node[scale=0.5] at (2.57,2.05) [red] {$\frac{1}{f_{\text{St}}}$};
  \draw[->] (2.19,2) -- (2.19,3);

        \put(6,4){\line(1,0){2}}
        \put(0.28,4){\vector(1,0){1.7}}
         \put(-2,4){\vector(1,0){1.7}}
        \put(8,1){\vector(-1,0){2}}
        \put(2,1){\line(-1,0){2}}
        \put(2,0){\framebox(4,2){$c^2_y(t)\dot{c}_y(t)$}}
        \node at (2.5,0) [below] {\footnotesize{static nonlinear function}};
        \put(8,1){\line(0,1){3}}
        \put(0,1){\vector(0,1){2.7}}
        \put(0.5,1){\makebox(1,1){$z(t)$}}
         \node[scale=0.6] at (5,3.1) {$\textbf{x}(t) = \left[ \begin{matrix} c_y(t) \\ \dot{c}_y(t) \end{matrix} \right]$};
        \put(-2,4){\makebox(1,1){$y(t)$}}
        \put(13,4){\makebox(1,1){$c_y(t)$}}
        \put(8,4){\circle*{0.3}}
        \put(8,4){\line(1,0){2}}
        \put(0,4){\circle{0.6}}
        \put(-0.5,3.5){\makebox(1,1){$+$}}
        \node at (0,2) [left] {$-$};
         \node at (0,2.8) [left] {$+$};
        
        \put(10,4){\line(0,1){1}}
        \put(10,5){\line(1,0){2}}
        \put(12,5){\line(0,-1){2}}
        \put(12,3){\line(-1,0){2}}
        \put(10,3){\line(0,1){1}}
        
        \put(12,4){\vector(1,0){2}}
        
         \node[scale=0.6] at (6.85,2.5) {$[1 \quad 0]~\textbf{x}(t)$};

\end{tikzpicture}
\end{center}
\vspace{-0.5cm}
\caption{Schematic representation of the Van der Pol equation (Eq.~\eqref{e:VdP2}) as a nonlinear feedback loop.}
\label{f:feedback_scheme}
\end{figure}
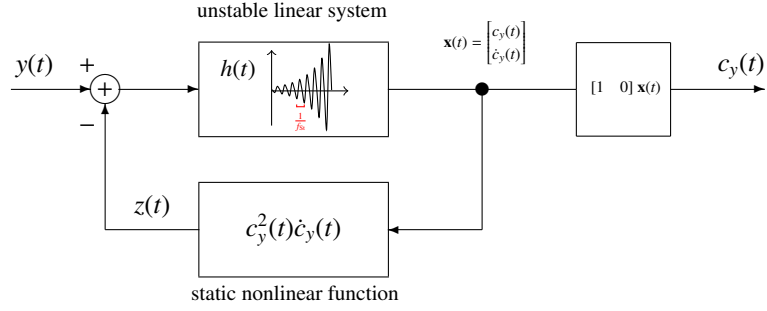
This is typically represented as a feedback loop, containing a linear dynamic system in the feedforward (left hand side of Eq.~\eqref{e:VdP2}) and a feedback path containing the nonlinear term. The latter is drawn schematically in Fig.~\ref{f:feedback_scheme}. Reading the loop from the input, $y(t)$, to the output, $c_y(t)$, it is interpreted as follows:
\begin{itemize}
\item The input is applied to a linear second-order system (left hand side of Eq.~\eqref{e:VdP2}). Since it has a negative damping term we have that the origin ($c_y(t) = 0$, $\dot{c}_y(t) = 0$) is an unstable fixed point \cite{dimitriadis2017}. This is illustrated by an exponentially growing impulse response, oscillating at the natural frequency of the linear system, i.e.\ $f_{\text{St}}$.
\item Since the impulse response does not decay to zero, a nonzero initial condition will result in an output of the linear system, oscillating at the natural frequency, i.e.\ $f_{\text{St}}$, and with an increasing amplitude. When additionally an input signal with a frequency $f_{\text{ex}}$ is applied, both $f_{\text{ex}}$ and $f_{\text{St}}$ will appear at the output of the linear system.
\item For convenience $c_y(t)$ and $\dot{c}_y(t)$ are stacked in a state vector $\textbf{x}(t)$. These internal states are fed back through the static nonlinear function $c^2_y(t)\dot{c}_y(t)$ and subtracted from the input.
\item The output is then defined as the first state variable.
\end{itemize}
If we consider the signals to be cosines and focus on their frequency content (removing phase and amplitude information from the expressions) we have that the input is described by
\begin{equation}
y(t) =\left[(e^{j\omega_{\text{ex}} t}+e^{-j\omega_{\text{ex}} t})\right]. \label{e:cy_cubic1}
\end{equation}
Ignoring phase and amplitude information (removing the effect of the derivative operator), the nonlinear feedback acts as a cubic function of a signal $x(t)$ with
\begin{equation}
x(t) = \left[(e^{j\omega_{\text{ex}} t}+e^{-j\omega_{\text{ex}} t})+(e^{j\Omega_{\text{St}} t}+e^{-j\Omega_{\text{St}} t})\right] \label{e:cy_cubic2},
\end{equation}
resulting in the signal
\begin{center}
\begin{multline}
z(t) = \left[(e^{j\omega_{\text{ex}} t}+e^{-j\omega_{\text{ex}} t})+(e^{j\Omega_{\text{St}} t}+e^{-j\Omega_{\text{St}} t})\right] \\ \times \left[(e^{j\omega_{\text{ex}} t}+e^{-j\omega_{\text{ex}} t})+(e^{j\Omega_{\text{St}} t}+e^{-j\Omega_{\text{St}} t})\right]\\ \quad \quad \times \left[(e^{j\omega_{\text{ex}} t}+e^{-j\omega_{\text{ex}} t})+(e^{j\Omega_{\text{St}} t}+e^{-j\Omega_{\text{St}} t})\right], \label{e:cy_cubic3}
\end{multline}
\end{center}}
with $\omega_{\text{ex}}=2\pi f_{\text{ex}}$ and $\Omega_{\text{St}}=2\pi f_{\text{St}}$. Expanding Eq.~\eqref{e:cy_cubic3} results in the combinatorial possibilities of summing three of the four frequencies $\{-f_{\text{St}},f_{\text{St}},-f_{\text{ex}},f_{\text{ex}}\}$,

\begin{equation}
\label{e:cubic_combi}
\left[ \begin{matrix} 
+f_{\text{St}} +f_{\text{St}} +f_{\text{ex}} \\
+f_{\text{St}} +f_{\text{St}} -f_{\text{ex}} \\
+f_{\text{St}} -f_{\text{St}} +f_{\text{ex}} \\
+f_{\text{St}} -f_{\text{St}} -f_{\text{ex}} \\
-f_{\text{St}} +f_{\text{St}} +f_{\text{ex}} \\
-f_{\text{St}} +f_{\text{St}} -f_{\text{ex}} \\
-f_{\text{St}} -f_{\text{St}} +f_{\text{ex}} \\
-f_{\text{St}} -f_{\text{St}} -f_{\text{ex}} \\
 \end{matrix} \right] = \left[\begin{matrix} 
 +2f_{\text{St}}+f_{\text{ex}}\\ 
 +2f_{\text{St}}-f_{\text{ex}}\\ 
 +f_{\text{ex}} \\ 
 -f_{\text{ex}} \\ 
 +f_{\text{ex}}\\
 -f_{\text{ex}} \\ 
 -2f_{\text{St}}+f_{\text{ex}} \\ 
  -2f_{\text{St}}-f_{\text{ex}}\\
    \end{matrix}\right],
\end{equation}
from which we can see that output power will be generated at $2f_{\text{St}}-f_{\text{ex}}$\footnote{It is worth noticing that, provided multiple iterations of the feedback loop are travelled, also a quadratic function would return a contribution at $2f_{\text{St}}-f_{\text{ex}}$. Having a quadratic function would however result in a clear presence of even nonlinear distortions in the FAST analysis of Section \ref{ss:FAST}.}.

\rev{It is hence shown that feature IV can be a product of having a nonlinear system with internal feedback which results in an interaction between input frequencies and the vortex related frequency $f_{\text{St}}$.

The analysis of the Van der Pol structure shows that in order to admit features as the ones observed during the sine-sweep experiment (Fig.~\ref{f:cy_0p20_spectro}) a model structure allowing for internal feedback and (at least) a third order nonlinearity is required.}

\subsection{Characterising the nonlinear behaviour using the FAST-approach}
\label{ss:FAST}

The objective of the FAST-approach is to characterise the type of nonlinearity \cite{pintelon2001,schoukens2016}. The method distinguishes even from odd nonlinear behaviour. Knowledge of the latter can be directly implemented when constructing an appropriate model, as will be shown in Section \ref{s:modelling}.

The approach exploits the property of nonlinear systems to generate output power at unexcited frequencies, similar to the features which were observed and discussed in Section \ref{ss:SWS}. The method relies on the usage of a specifically constructed excitation signal called random-phase random-odd multisines \cite{pintelon2001}.  A random-phase random-odd multisine is a sum of harmonically related sines with randomly selected phases,
\begin{equation}
\label{e:multisine}
y(t) = \sum_{n=1}^N A_n \sin\left[2\pi (2n-1)f_0t+\phi_n\right],
\end{equation}
where $A_n$ enables to select an appropriate amplitude spectrum and $\phi_n$ is a random phase, drawn from a uniform distribution between $[0,2\pi)$. The signal is constructed such that it covers the frequency band of interest, thereby only exciting odd multiples of the base frequency, $f_0$. In the case of a random-odd multisine, $A_n$ is selected in such a way that for every bin of 4 odd frequency lines, one frequency line is not excited, i.e.\ $A_n=0$. The odd lines which are excluded are called detection lines and the position of these lines within a bin of 4 odd lines is chosen randomly. Generally the remaining odd excited lines are given an equal amplitude, resulting in a flat amplitude spectrum. 

Distinguishing odd from even nonlinear behaviour is done by studying the amplitude spectrum of \CYc. 

\subsection{Illustration of the FAST-approach}

\begin{figure}
\begin{center}
\begin{subfigure}[b]{\textwidth}
\begin{center}
\includegraphics[width=0.8\textwidth]{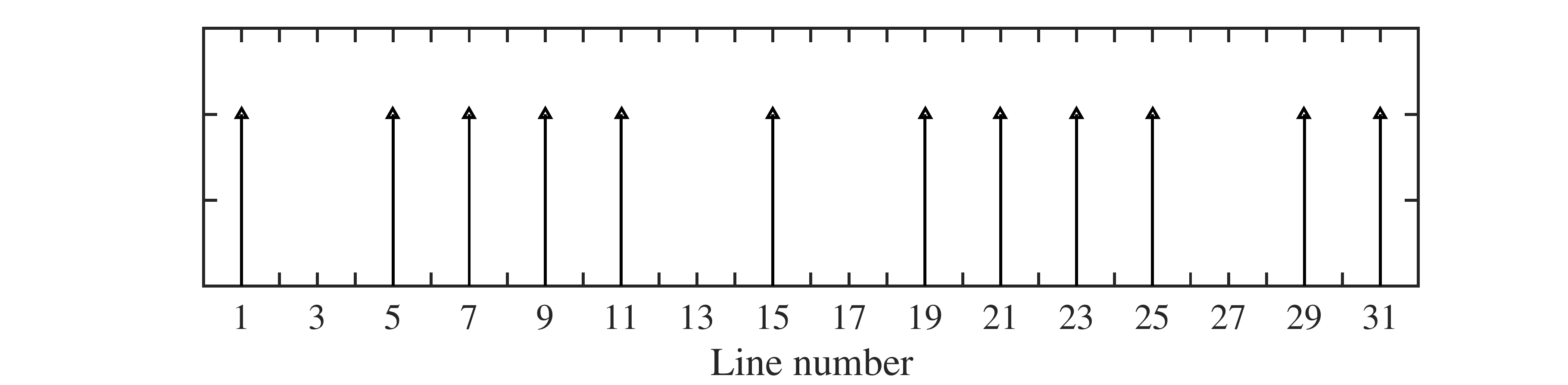}
\caption{}
\label{f:multisine_input}
\end{center}
\end{subfigure}
\begin{subfigure}[b]{\textwidth}
\begin{center}
\includegraphics[width=0.8\textwidth]{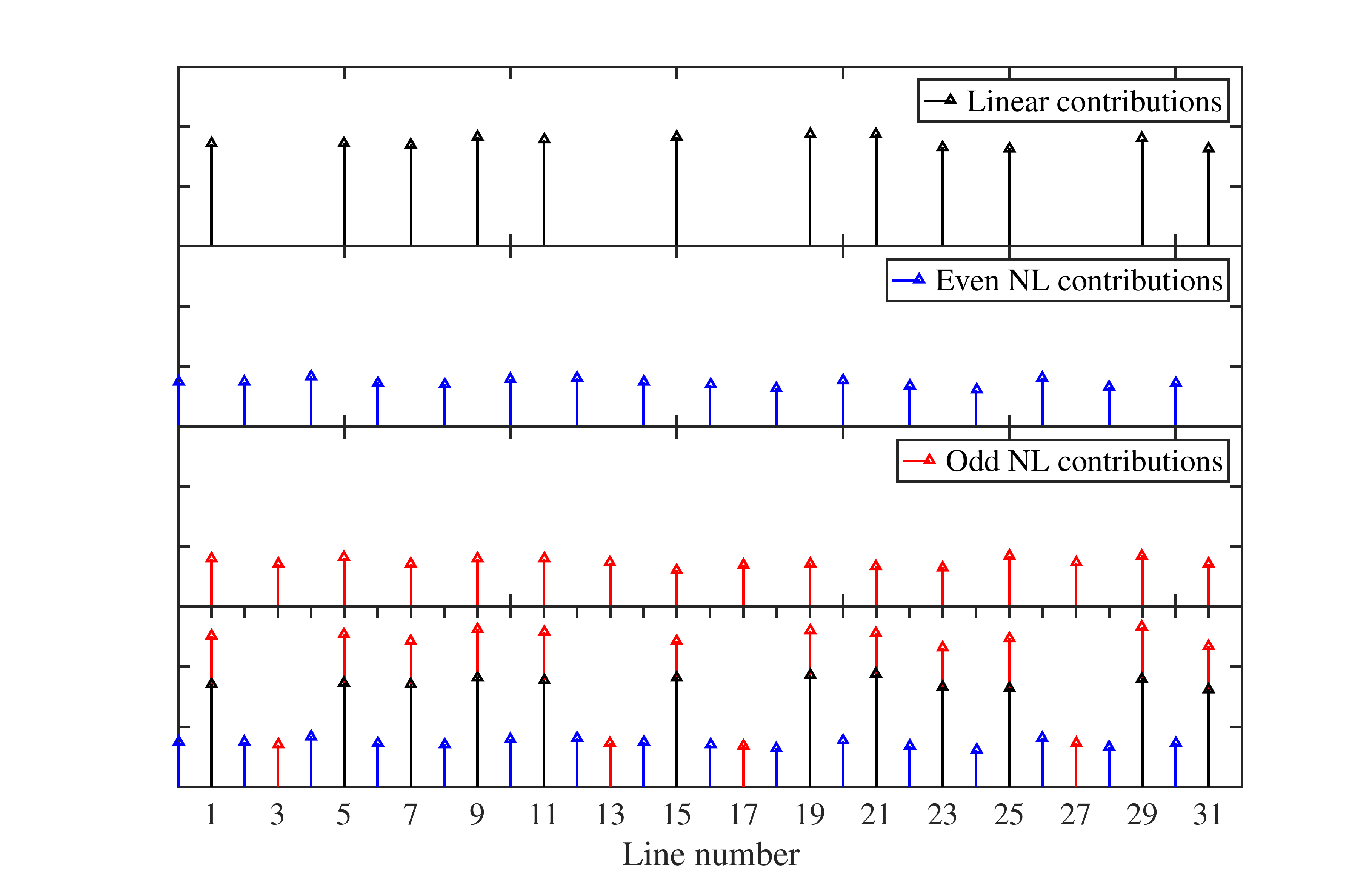}
\caption{}
\end{center}
\end{subfigure}
\caption{Example of the DFT spectra of random-phase random-odd multisine (a) and the different output contributions that can follow from it when it passes through a nonlinear system.}
\label{f:FAST_example}
\end{center}
\end{figure}

The output resulting from a random-odd random-phase multisine can be decomposed into linear, odd nonlinear and even nonlinear contributions, based on the frequency lines where power is present. Generally the following holds:
\begin{itemize}
\item Linear transformations only result in a scaling of the applied amplitude (only the magnitude is studied). Hence the frequency of excitation is preserved. Therefore all linear contributions will only be present at the excited lines. 
\item Since the even lines are unexcited in an odd \rev{multisine}, no linear contributions will be present here. To arrive at an even line, an even nonlinear function is needed, combining an even number of excited lines (similar to the product given by Eq.~\eqref{e:cy_cubic3}). The even lines therefore directly serve as quantifier for the amount of even nonlinear behaviour present in the system.
\item Similarly, at the odd unexcited lines (detection lines), there can only be contributions from odd nonlinearities.
\item  Consequently, at the excited odd lines there can be both, odd nonlinear contributions and linear contributions. Both contributions can be separated from each other by extrapolating the odd nonlinear contributions on the unexcited odd lines to the excited lines and subtracting them from the total contribution at the excited odd lines.
\end{itemize}
In Fig.~\ref{f:FAST_example} the DFT of an example random-odd  random-phase multisine input signal is shown (panel a). Below (panel b) the different contributions are plotted, first separately, and then summed together representing the case where the input signal would have passed through a nonlinear system containing both odd and even nonlinearities.

\subsection{Applying the FAST-approach to the fluid system}
\begin{figure}
\begin{center}
\begin{subfigure}[b]{\textwidth}
\begin{center}
\includegraphics[width=\textwidth]{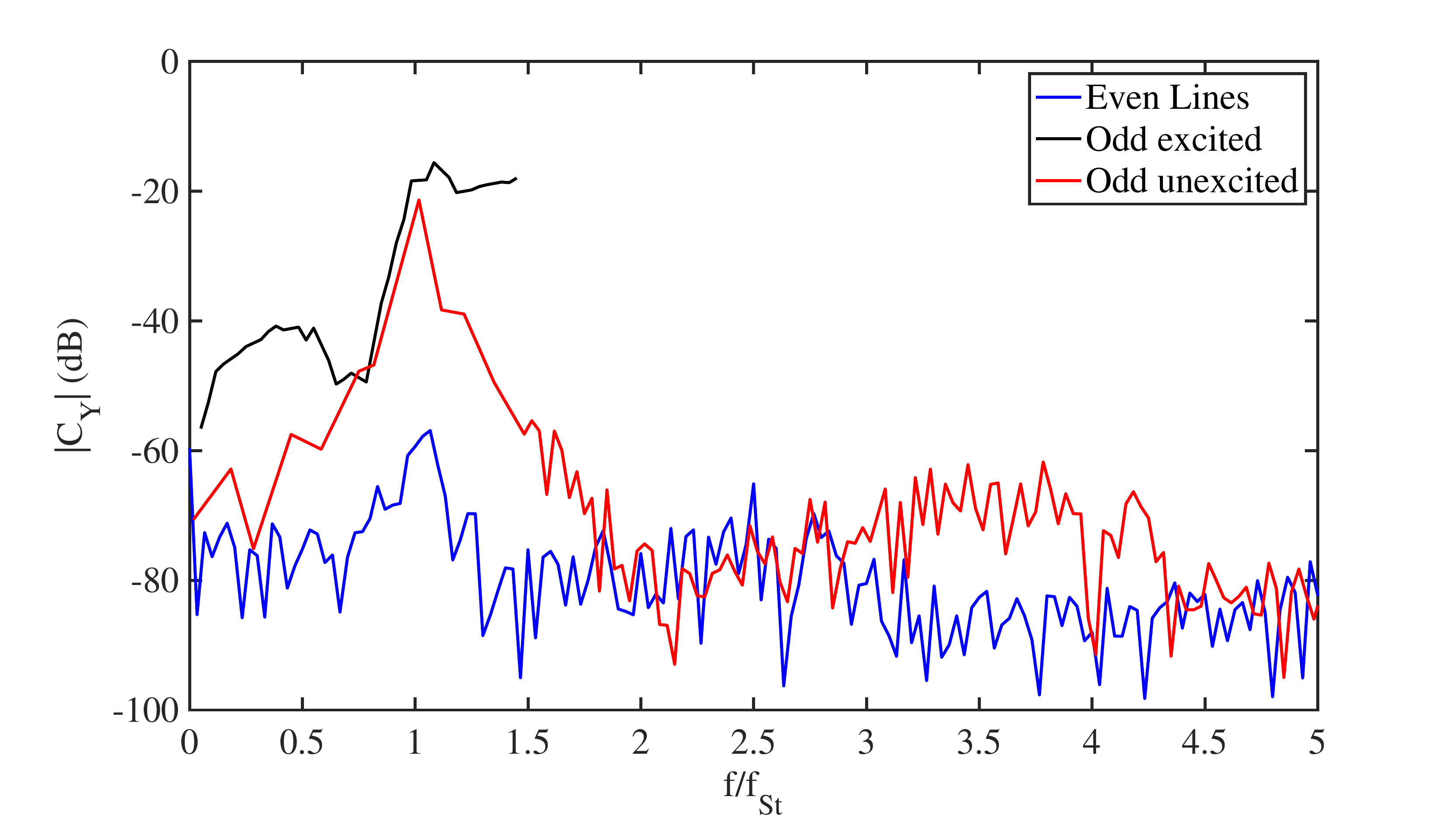}
\caption{}
\end{center}
\end{subfigure}
\begin{subfigure}[b]{\textwidth}
\begin{center}
\includegraphics[width=\textwidth]{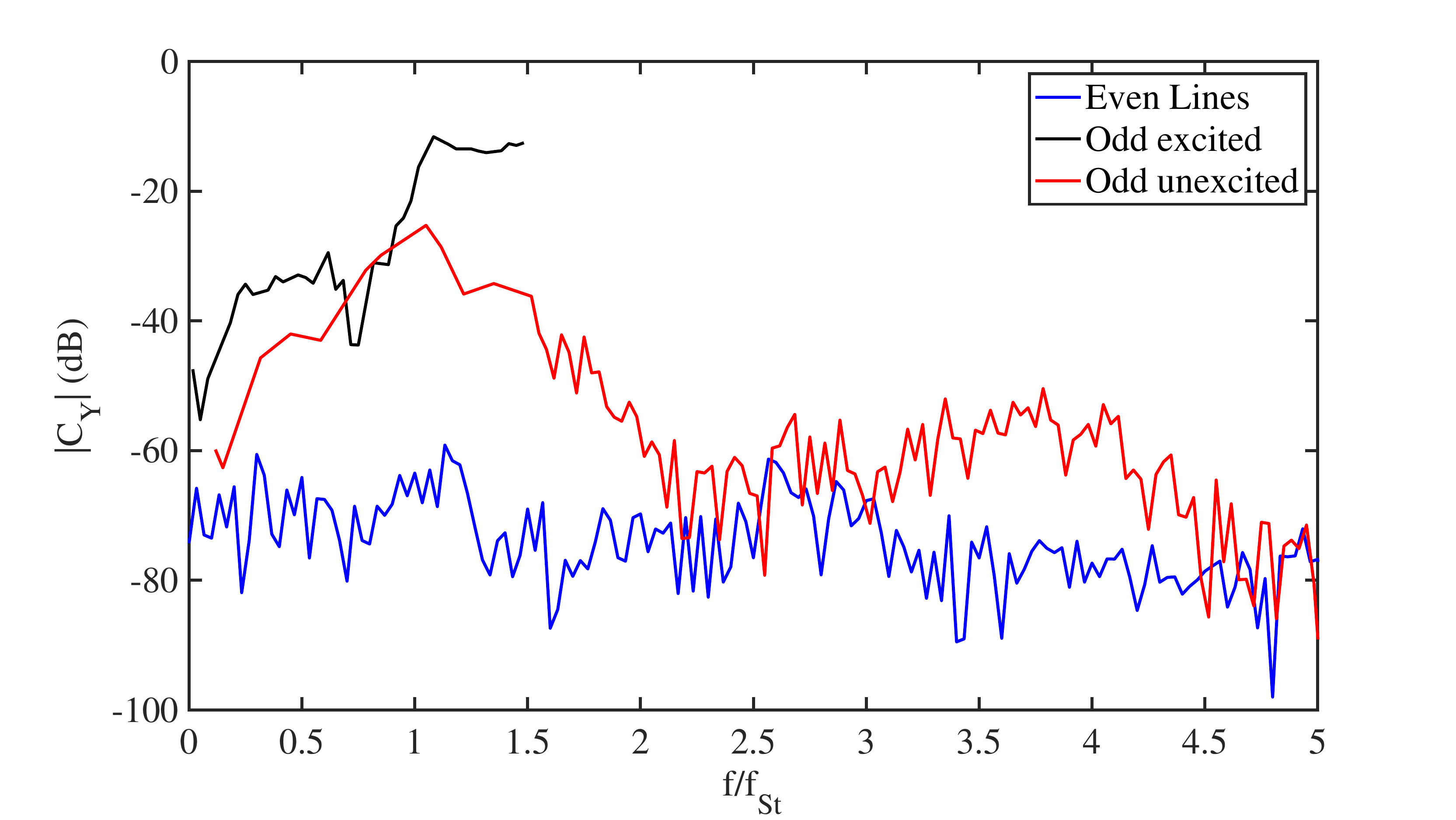}
\caption{}
\end{center}
\end{subfigure}
\caption{Amplitude spectra of \CY from random-odd multisine excitation. (a) Amplitude spectrum of the FAST-test at $A/D=0.15$. (b) Amplitude spectrum of the FAST-test at $A/D=0.30$. For clarity no subharmonic levels are shown.}
\label{f:FAST}
\end{center}
\end{figure}

In this section the FAST-approach is used to characterise the nonlinear behaviour of the fluid system. 
Fig.~\ref{f:FAST} shows the DFT output spectra at 2 excitation amplitude levels, $A/D=0.15$ and $0.30$ and a frequency band $f_{\text{ex}}/f_{\text{St}}= 0 - 1.5$. \rev{The random-phase random-odd multisine was constructed with a base frequency of $f_0 = 0.05$ Hz and sampled at $f_s = 50$ Hz. The averaged output, measured over 12 periods of the base frequency and after removal of transients, was used.} 

We observe that at both amplitudes the odd nonlinear contributions dominate. Again the presence of a third harmonic is visible, especially related to the high frequency range of excitation, $f_{\text{ex}}>f_{\text{St}}$, resulting in a bump around $f/f_{\text{St}}>3$. Even contributions are hardly present  (40 dB lower than the main contributions).

Due to a shift of the vortex shedding frequency, observed for oscillating cylinders, the frequency-domain plot is distorted by leakage. To reduce the impact a Hanning window was applied. 

Important to notice is that the magnitude of the odd nonlinear contributions reaches as high as the linear contributions, in the neighbourhood of $f_{\text{St}}$. The considerable amount of nonlinearity indicates that a model, aiming at accurately  describing the relationship between $y(t)$ and $c_y(t)$, will necessarily have to be nonlinear.




\paragraph{To summarise: from analysing the output spectrum of a swept sine imposed motion experiment it was found that output power is present at harmonics of the combined frequencies $f_{\text{ex}}$ and $f_{\text{St}}$. This is evidence that the relationship between the imposed motion $y(t)$ and the transverse force coefficient $c_y(t)$ can be considered as a feedback loop, containing a nonlinear block. Using random-phase random-odd multisine signals as excitation, the nonlinearity in the system was characterised as dominantly odd and of a considerable level. This insight is used in constructing a nonlinear model in Section \ref{s:modelling}}

\section{Modelling the transverse force coefficient}
\label{s:modelling}

A model class which inherently incorporates feedback is the state-space class of models \cite{kailath1980}. \rev{A state-space model is a formulation of a higher order differential (or difference) equation as a set of first order differential (or difference) equations. The intermediate state variables, denoted by $\textbf{x}$, convey the memory of the system while linking the instantaneous input to the instantaneous output.} In a nonlinear version the classical state equation (Eq.~\eqref{e:PNLSS2_a}) and the output equation (Eq.~\eqref{e:PNLSS2_b}) are extended with nonlinear functions \cite{tiels2016}. In this work we use polynomial expansions \cite{paduart2010}. This allows for an easy selection between odd and even nonlinear terms. Expanded, the polynomials are written as the matrix products of two sets of coefficients, matrices \textbf{E} and \textbf{F}, and two sets of monomial basis functions $\bm{\eta}$ and $\bm{\zeta}$, which are monomials in the states $\textbf{x}$ and the input $y$. \rev{It is crucial that the state variables are used in $\bm{\zeta}$ such that nonlinear feedback can be modelled.} This results in a polynomial nonlinear state-space (PNLSS) model
\begin{subequations} \label{e:PNLSS2}
    \begin{empheq}[left={\empheqlbrace\,}]{align}
      & \textbf{x}(k+1)=\textbf{A}\textbf{x}(k)+\textbf{b}y(k)+\textbf{E}\bm{\zeta}(k) \label{e:PNLSS2_a} \\ 
      & \CYc(k)=\textbf{c}^{\text{T}}\textbf{x}(k)+dy(k)+\textbf{f}^{\text{T}}\bm{\eta}(k) \label{e:PNLSS2_b}
    \end{empheq}
\end{subequations}
where $k$ denotes the time index. \rev{Boldfaced upper case letters denote matrices and boldfaced lower case letters are vectors. Given that the considered system is characterised by a single input, the displacement $y(t)$ and a single output, $c_y(t)$, the matrices have the following dimensions: $\textbf{A} \in \mathbb{R}^{n \times n}$ with $n$ the size of the state vector, $\textbf{b} \in \mathbb{R}^{n}$, $\textbf{c} \in \mathbb{R}^n$ and $d$ is a scalar.} If the data is sampled with a sampling period $T_s$, $y(k)$ is the sample of the input at time instant $t = kT_s$. The vectors $\bm{\zeta}(k) \in \mathbb{R}^{n_\zeta}$ and $\bm{\eta}(k) \in \mathbb{R}^{n_\eta}$ contain the nonlinear terms of the polynomial (monomials). The monomials in $\bm{\zeta}(k)$ and $\bm{\eta}(k)$ are formed as cross products between the input (imposed motion $y(t)$) and the state variables ($\textbf{x}(t)$) raised to a chosen degree. Given the insight gained from the FAST-approach in Section \ref{ss:FAST}, only odd degrees are chosen.

One way of generating such a set is by constructing all the combinatorial cross products between states and input of which the total degree, i.e.\ the sum of the degrees of the individual factors, satisfies a given value. An example of an element in $\zeta(k)$ is given by:
\begin{equation}
\label{e:PNLSS3}
\zeta_{g,h_1,...,h_{n}}(t)=y^g(k)\prod_{i=1}^{n} x_i^{h_i}(k)
\end{equation}
with the total degree of the monomial satisfying the condition: $g+\sum_{i=1}^{n} h_i \in \left\{0,3,5,...,p \right\}$ with $[g,h_i] \in \mathbb{N}$ and $n$ being the order of the underlying linear model. In practice $p=7$ was chosen after scanning over increasing odd degrees and selecting the best one on the basis of validation data. Odd polynomial terms have commonly been proposed when modelling fluid forces, \rev{e.g.\ the Van der Pol equation (Eq.~\eqref{e:VdP})}, frequently used in wake-oscillator models (see \cite{gabbai2004} and the references cited therein), contains a third order nonlinear damping term. Also the Parkinson's galloping model contains a $7^{th}$ order polynomial term \cite{dimitriadis2017}.

\subsection{The approach}
The \rev{data-driven} identification procedure consists of 3 major steps \cite{paduart2010}:
\begin{enumerate}
\item First, a nonparametric linear model is extracted from data. \rev{This is of the form of a nonparametric frequency response function.}
\item Next, linear system identification techniques are used to obtain a linear parametric estimate of the model.
\item The parametric estimate provides initial values for the \rev{$\textbf{A}$, $\textbf{b}$, $\textbf{c}$ and $d$ coefficients; $\textbf{E}$ and $\textbf{f}$} are set to zero. The final estimate of the full nonlinear model is then obtained by \rev{minimising a least-squares cost function quantifying the distance between the modelled output and the true output. The optimisation is computed using a Levenberg-Marquardt algorithm.}
\end{enumerate}

In this work, the linear model which is used as initialisation is calculated as the mean of the \emph{best linear approximation} (BLA), averaged over the amplitude levels starting at $A/D=0.05$ up to $A/D=0.30$ in steps of $A/D=0.05$. The BLA is a linear approximation of a nonlinear system which can be found by applying random-phase multisine excitation signals as the input to the system \cite{Dhaene2005}. \rev{Proper initialisation is vital to ensure convergence to a good local minimum of the cost function. }Alternative linear initialisation options were described in \cite{decuyper_phd_2017}.

From this nonparametric estimate and the sample covariance, which are obtained directly from the data, a parametric linear estimate \rev{i.e.\ a linear state-space model}, is identified. This was done using the \emph{frequency domain identification toolbox} (FDIDENT) in MATLAB \cite{FDIDENT}. A parametric \rev{finite impulse response} model of fifth order (5 state variables) was found adequate \rev{from a model scan}.

The final nonlinear optimisation step is performed using a Levenberg-Marquardt algorithm \cite{levenberg1944,fletcher1981}. A detailed discussion on extracting the BLA from data of \CY can be found in \cite{decuyper2018}.

\subsection{Tuning the model to the data}

In \cite{decuyper2018} it was shown that data of swept sine signals which cover the frequency and amplitude range of interest, can be used as a training data set for the nonlinear model. \revrev{Even though sine sweeps are non-stationary signals, they are rich enough to cover a wide region of phase-space, resulting in accurate reproduction of stationary single harmonics as well (see Section \ref{ss:val}).} The data used here consists of the concatenation of swept sine signals sweeping over the frequency band $f_{\text{ex}}/f_{\text{St}}=0-1.5$ at the discrete amplitude levels from $A/D=0.05$ up to $A/D=0.30$ in steps of $0.025$. This constitutes the exact same training data sequence as in \cite{decuyper2018}. Hence the benefit of using a tailored nonlinear state-space structure, tailored given the knowledge of the FAST-test, can be studied.

\begin{figure}
\begin{center}
\includegraphics[width=\textwidth]{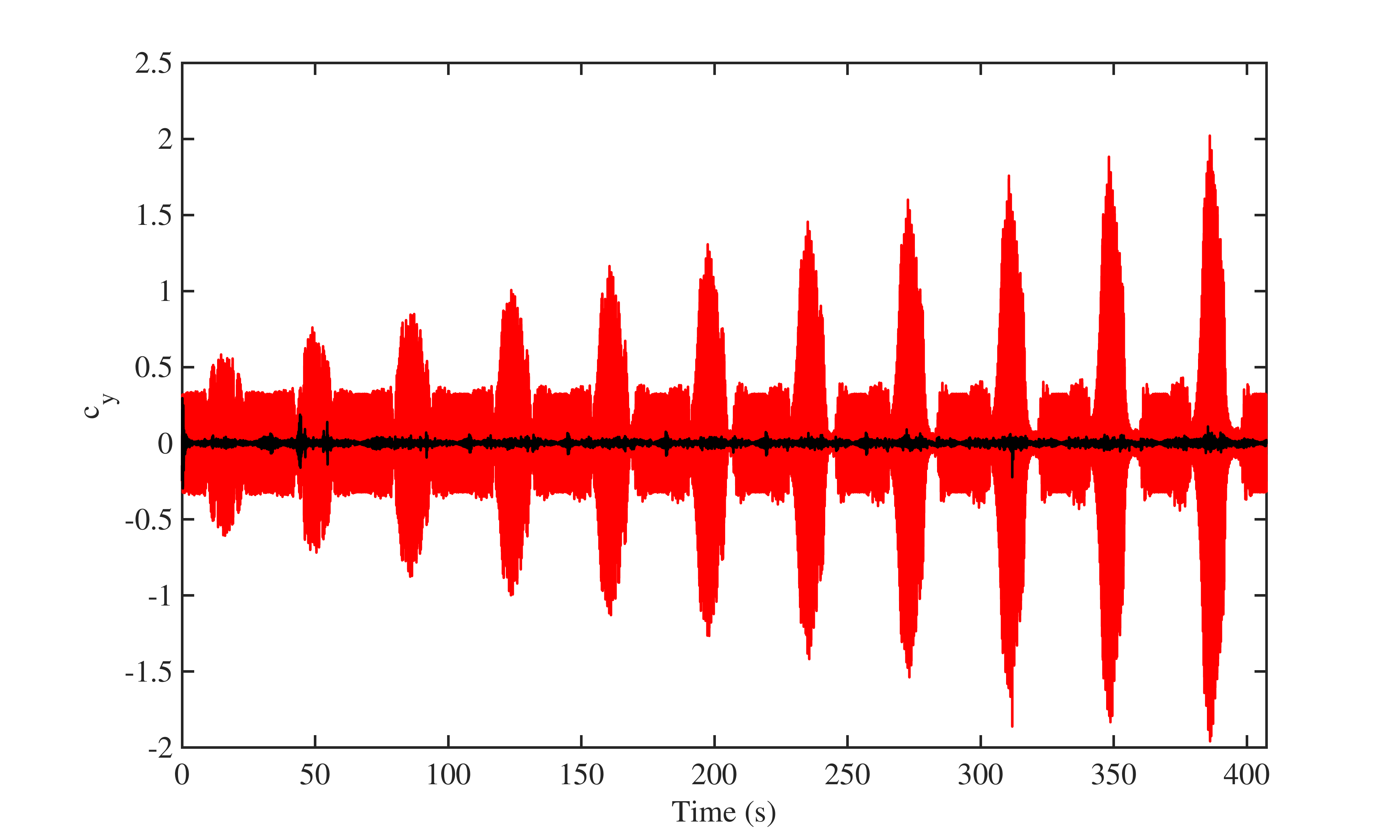}
\caption{Error (in black) on the simulated training set of the best performing estimated PNLSS model, true output in red.}
\label{f:model_est_error}
\end{center}
\end{figure}

As quantifier of the accuracy of the fit to the data, the relative root-mean-square error is used:
\begin{equation}
\label{e:e_rms}
 e_{\text{rms}} = \frac{\sqrt{\frac{1}{N}\sum_{k=1}^N (c_y(k)-c_{y,\text{sim}}(k))^2}}{\sqrt{\frac{1}{N}\sum_{k=1}^N c_y(k)^2}},
 \end{equation}
with $N$ the number of data points, $c_y$ the true output and $c_{y,\text{sim}}$ the output simulated by the PNLSS model. After optimisation, the nonlinear  model yields a low relative r.m.s.\ simulation error of only $e_{\text{rms}}=0.04$. The accurate fit is depicted in Fig.~\ref{f:model_est_error} where the output of the training set is shown together with the error on the simulated data. The error is calculated as the difference between the training data and the output of the PNLSS model at every time step. 

\subsection{Validation results}
\label{ss:val}

\begin{figure}
\begin{center}
\begin{subfigure}[b]{\textwidth}
\begin{center}
\includegraphics[width=\textwidth]{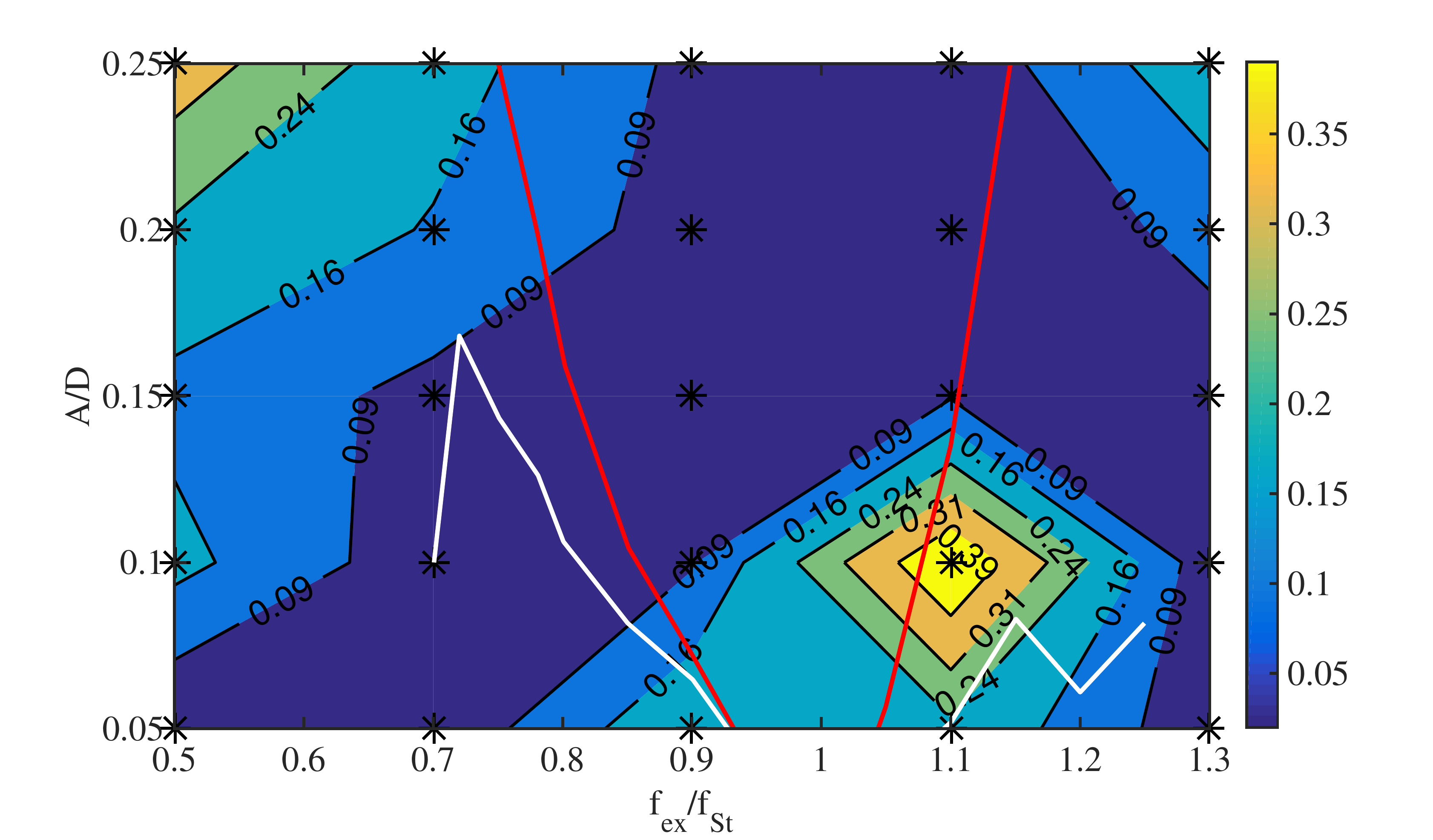}
\caption{}
\label{f:contour_erms1}
\end{center}
\end{subfigure}
\begin{subfigure}[b]{\textwidth}
\begin{center}
\includegraphics[width=\textwidth]{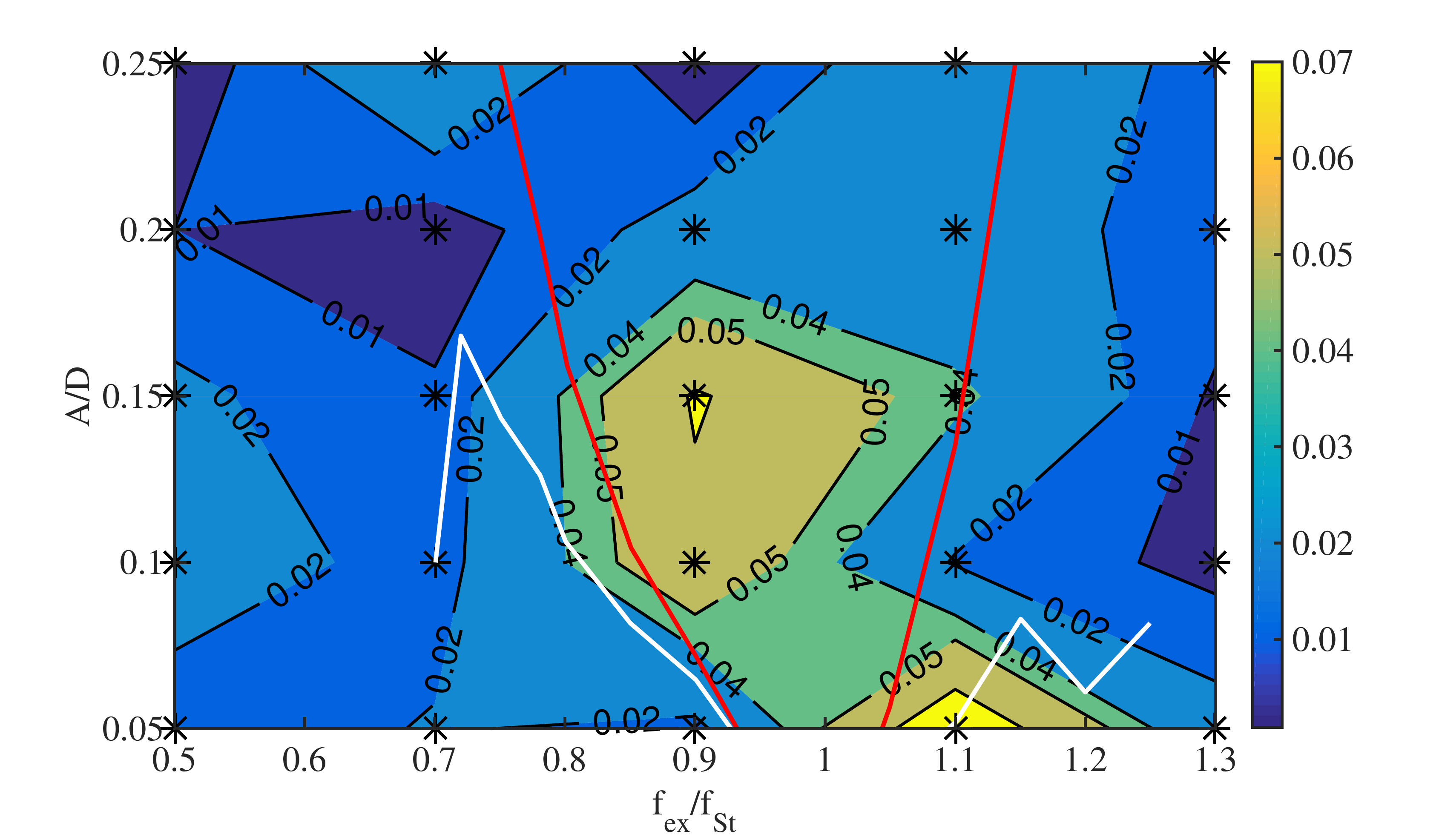}
\caption{}
\label{f:relMaxAmp}
\end{center}
\end{subfigure}
\caption{\color{black}Single harmonic sine validation results at 25 test locations throughout the frequency-amplitude plane of imposed motion. Red indicates the boundaries of the lock-in region (based on \cite{kumar2016}). (a) Relative root-mean-squared error ($e_{\text{rms}}$). (b) Relative error on the maximum amplitude level.}
\end{center}
\end{figure}

To assess the model quality, it is subjected to a validation test. The test consists of simulating the output to a single harmonic sine input signal. A total number of 25 validation time series, acquired at various locations throughout the frequency-amplitude plane of imposed motion are tested. The results are represented by a contour plot of the corresponding $e_{\text{rms}}$ values in Fig.~\ref{f:contour_erms1}. Red and white lines indicate the boundaries of the lock-in and so-called transition regions respectively \cite{kumar2016}. The model performs well on the majority of the validation points with errors below 5\%. These are errors which are in the same order of magnitude as the uncertainty found on CFD models in literature (see Table \ref{t:val_stat_3Hz}). 

\revrev{Two regions with elevated errors are found. One is located at the point $f_{\text{ex}}/f_{\text{St}}=1.1$ and $A/D=0.1$. The reason is attributed to the usage of a different type of excitation signal for training (swept sine) and for validation (single sine). Although many of the features were accurately captured from the swept sine training, slight differences in the system's response may be present, e.g.\ the exact location of the boundary of the lock-in range is known to depend on the type of excitation \cite{decuyper_phd_2017}. This local mismatch between training data and validation data results in an error. To improve the results, the shortcomings of the training data could be tackled during a second tuning step where additional data are added for the poorly captured regimes (cfr.~bagging and boosting in machine learning). A second region is found in the upper left corner of the domain. In this region the wake is unsynchronised, therefore the force signal will contain two frequency components, one at the imposed oscillation frequency and another at the Strouhal frequency. Given that there is no synchronisation, an arbitrary phase angle exists between both components. Arbitrary since it is only dictated by the time instant on which the input was initiated and the state of the wake at this particular instance. Phase shifts on the Strouhal component are therefore to be expected when validating in this region. Moreover, the quality measure which is used here, i.e.\ the relative root-mean-squared error, is very sensitive for phase errors (due to its point-wise error computation). We can additionally assess the performance by studying a relative maximum amplitude error criterion. This might well be one of the most crucial force features for physical structures. The results are shown in Fig.~\ref{f:relMaxAmp}. It can be observed that the relative error on the maximum of the amplitude is generally below 2\% and does not exceed 5\% considering all cases.}

\begin{figure}
\begin{center}
\includegraphics[width=\textwidth]{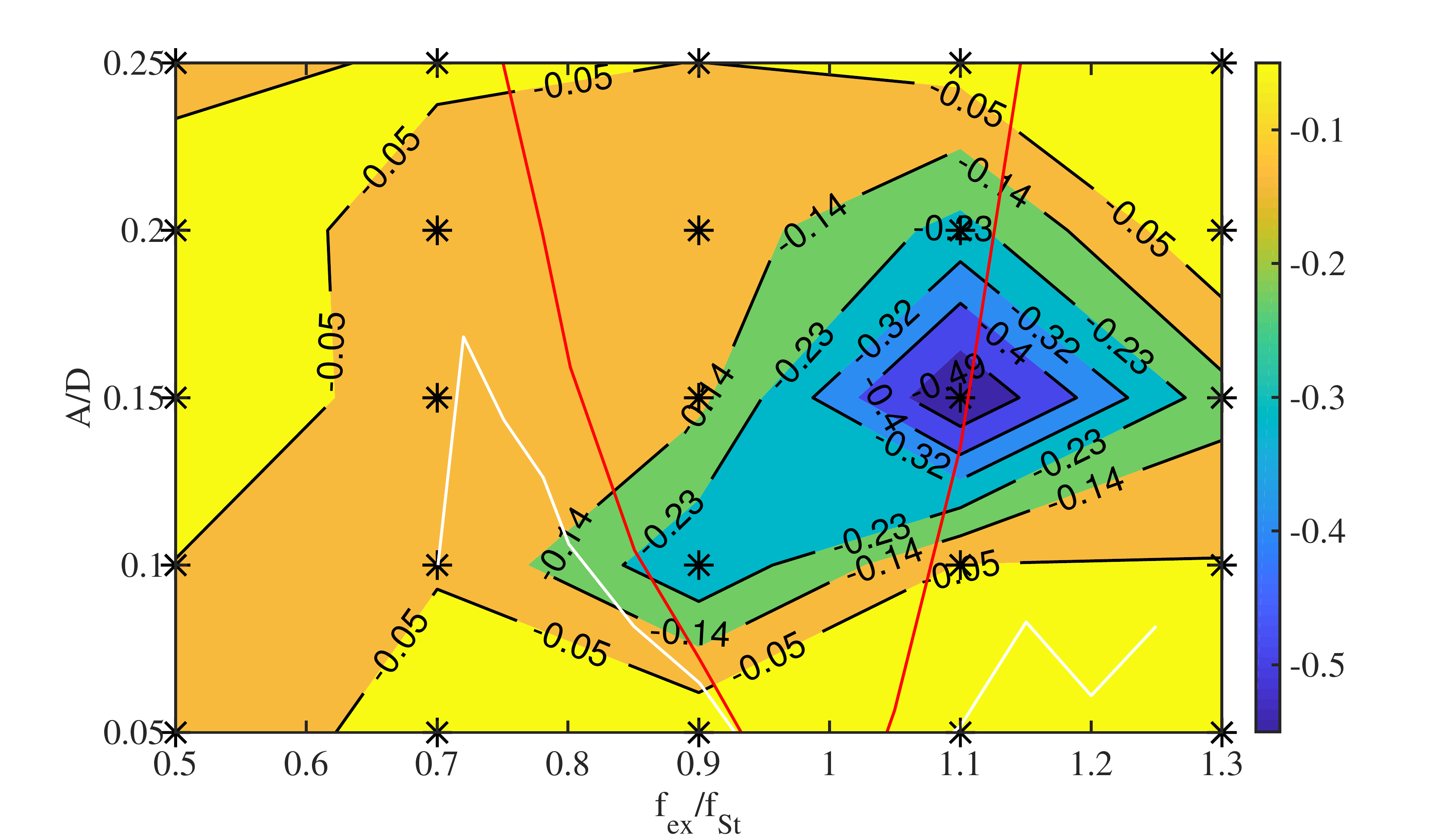}
\caption{Reduction of the $e_{\text{rms}}$ of the present study compared to the results reported in \cite{decuyper2018}.}
\label{f:erms_reduction}
\end{center}
\end{figure}

Fig.~\ref{f:erms_reduction} shows to what extent the error is reduced when compared to the results of \cite{decuyper2018}. It can be seen that the model with tailored basis functions (present study) outperforms the generic model on all validation points. Error reductions as much as $-50\%$ are obtained.

Overall a mean $e_{\text{rms}}$ value of $0.11$ is found for the present study as opposed to $0.20$ for the generic model structure. Since in that case no knowledge of the nonlinearity was used, also the even nonlinear monomials were used (degrees 0 up and till 5). \rev{This resulted in a model with 2594 degrees-of-freedom\footnote{\color{black}In this context, degrees-of-freedom indicate the number of independent parameters. Not to be mistaken with the degrees-of-freedom of the system which is only 1 in this case (single-output model).} as opposed to 1396 for the model of the present study.}  Including even nonlinear basis functions to model a dominantly odd nonlinear system unnecessarily increases the number of parameters (degrees of freedom) of the estimation process. This hinders the optimisation with as a result a less accurate model. \revrev{Not withstanding the relatively large number of parameters a significant time reduction is achieved by the reduced-order model, when compared to CFD. The model does not require time integration. Instead, a number of polynomial evaluations and matrix products are computed. This enables very quick computation of the output, e.g.\ simulating a time record of 20 s (real time) has a computing time below 1 s while the CFD simulation time was over 40 hours (on a 8 x 2.4 GHz machine). This is a speed-up of approximately 144000.}

\rev{
\subsection{Future potential of the PNLSS approach}
In order to bridge the gap between the current study and a true slender body undergoing VIV one has to extend the domain to the third dimension. Also in this context a PNLSS model could offer a solution. It is important to notice that in its most generic form, a PNLSS model allows for multiple inputs and multiple outputs:
\begin{subequations} \label{e:PNLSS3}
    \begin{empheq}[left={\empheqlbrace\,}]{align}
      & \textbf{x}(k+1)=\textbf{A}\textbf{x}(k)+\textbf{B}\textbf{y}(k)+\textbf{E}\bm{\zeta}(k) \label{e:PNLSS3_a} \\ 
      & \textbf{c}_y(k)=\textbf{C}\textbf{x}(k)+\textbf{D}\textbf{y}(k)+\textbf{F}\bm{\eta}(k) \label{e:PNLSS3_b},
    \end{empheq}
\end{subequations}
with the matrices having the following dimensions; $\textbf{A} \in \mathbb{R}^{n \times n}$, with $n$ the size of the state vector, $\textbf{B} \in \mathbb{R}^{n \times m}$, with $m$ the number of inputs, $\textbf{C} \in \mathbb{R}^{p \times n}$ with $p$ the number of outputs, $\textbf{D} \in \mathbb{R}^{p \times m}$, $\textbf{E} \in \mathbb{R}^{n \times n_{\zeta}}$ and $\textbf{F} \in \mathbb{R}^{p \times n_{\eta}}$.

Within this framework, a 3-dimensional structure could be modelled by a discrete set of points (also nodes) along the span. Each spanwise node is then assigned an input-output couple such that the input vector corresponds to the local displacement of the structure while the output vector describes the corresponding (local) lift coefficients. Via the intermediate layer of state variables, nonlinear interrelationships can be built between output variables, allowing to capture spanwise correlation and dynamics between nodes. 

\Rev{One might additionally want to consider a variable flow velocity ($U_{\infty}$) as opposed to the constant Reynolds number of the present study. Varying the Re-number corresponds to increasing the dimensions of the input space of the system. Also in this case the multiple input potential of the PNLSS models can be exploited. Designating $U_{\infty}$ to one of the inputs increases the input space of the model accordingly to that of the system.}
}

\section{Conclusion}
In this work an accurate nonlinear model relating the imposed motion of a circular cylinder in a uniform flow to the transverse force coefficient was obtained. A \rev{data-driven} black box modelling strategy was used. Data were acquired from 2-dimensional CFD simulations at a constant Re = 100. Using broadband excitation signals as imposed motion, a number of nonlinear features, characterising the fluid force, were observed and discussed. From a swept sine excitation it could be inferred that the fluid system behaves as a nonlinear feedback system. Using the so-called FAST-approach, the type of nonlinearity was investigated and found to be dominantly odd. This knowledge was used \rev{to select} an appropriate nonlinear model structure. The polynomial nonlinear state-space structure was selected and tuned to a training data set of swept sine input signals. The acquired model was validated on a large number of single sine validation experiments, corresponding to locations spread out over the frequency-amplitude plane of imposed motion. The model was found to outperform previous reported results on all validation sets. The mean relative r.m.s.\ error was reduced by half when the knowledge of the nonlinearity was explicitly used in constructing the model.

\section*{Acknowledgements}

This work was supported in part by the Fund for Scientific Research (FWO-Vlaanderen), the ERC advanced grant SNLSID, under contract 320378, the Swedish Research Council (VR) via the project NewLEADS -- New Directions in Learning Dynamical Systems (contract number: 621-2016-06079), and by the Swedish Foundation for Strategic Research (SSF) via the project ASSEMBLE (contract number: RIT15-0012). The authors would like to thank the reviewers for their valuable comments.

\appendix
\rev{
\section{Validation of the CFD data}
\label{s21:CFD_validation}

The CFD data are validated on the basis \rev{of} the following quantities of interest where arithmetic mean quantities are denoted with an over-bar and r.m.s.\ values with an asterisk:
\begin{itemize}
\item $-\overline{c}_{pb}$, base suction coefficient: the negative of the time-averaged pressure coefficient $c_p$ at the base of the cylinder ($x=0.5D, y=0$ for the stationary cylinder). As reference pressure the static pressure $p=0$ was used.
\item $\overline{c}_d$: time-averaged drag coefficient per unit span $c_d = \frac{F_x}{\rfrac{1}{2}\rho U_{\infty}^2}$ with $F_x$ the resultant force in the $x$-direction as indicated in Fig.~\ref{f:domain}.
\item $c_y^*$: r.m.s.\ \rev{value} of the transverse force coefficient per unit span $c_y = \frac{F_y}{\rfrac{1}{2}\rho U_{\infty}^2}$ with $F_y$ the resultant force in the $y$-direction as indicated in Fig.~\ref{f:domain}.
\item $\text{St}$: Strouhal number. Calculated on the basis of the frequency of \CYc. The frequency is obtained from analysing the \rev{spectrum calculated as the discrete Fourier transform of the signal.}
\item $\overline{\theta}_s$: Time-mean separation angle. The separation point is characterised by a zero wall vorticity along the cylinder surface \cite{qu2013,braza1986}. In practice vorticity is never exactly zero, therefore the point of minimum vorticity (in absolute value) is used. The angle is measured starting from the rear end of the cylinder.
\end{itemize}
For all quantities that involve time averaging, the number of periods used (of the slowest yet significant frequency component present in the signal) is reported. In all cases, transients were discarded based on visual inspection of the signal. The results are presented in Table \ref{t:val_stat_3Hz}, together with reference values from literature.

\begin{table}[h!]
\begin{center}
\caption{Validation results on stationary cylinder simulations at Re $=100$. Over-bars denote time averaging while an asterisk denotes r.m.s.\ values.}
\label{t:val_stat_3Hz}
\begin{tabular}{| l | c  c  c c c|}
\cline{2-6}
 \multicolumn{1}{c|}{} & $-\overline{c}_{pb}$  & $\overline{c}_d$ & $c_y^*$ & $\text{St}$ & $\overline{\theta}_s$ ($^{\circ}$)  \\
\cline{1-6}
Park et al., 1998 \cite{park1998} & 0.725 & 1.33 & 0.235 & 0.165 & - \\
Kravchenko and Moin, 1998 \cite{kravchenko1998} & 0.73 & 1.32 & 0.222 & 0.164 & -\\
Shi et al., 2004 \cite{shi2004} & - & 1.318 & - & 0.164 & - \\
Mittal, 2005 \cite{mittal2005} & - & 1.322 & 0.226 & 0.1644 & -\\
St\aa lberg et al., 2006 \cite{stalberg2006} & - & 1.32 & 0.233 & 0.166 & -\\
Posdziech et al., 2007 \cite{posdziech2007} & 0.709 & 1.325 & 0.228 & 0.1644 & -\\
Li et al., 2009 \cite{li2009} & 0.701 & 1.336 & - & 0.164 & -\\
Qu et al., 2013 \cite{qu2013} & 0.709 & 1.319 & 0.225 & 0.1648 & 118.0\\
Wu et al., 2004 \cite{wu2004} & - & - & -   & - & 117.02\\
Engelman et al., 1990 \cite{engelman1990} & - & 1.40 & 0.257 & 0.173 & -\\
Kang et al., 1999 \cite{kang1999} & - & 1.34 & 0.236 & 0.164 & - \\
Sharman et al., 2005 \cite{sharman2005} & 0.72 & 1.32 & 0.230 & 0.164 & -\\
Burbeau et al., 2002 \cite{burbeau2002} & - & 1.41 & 0.257 & 0.164 & -\\
Muralidharan et al., 2013 \cite{muralidharan2013} & - & 1.41 & 0.242 & 0.167 & -\\
Kim et al., 2001 \cite{kim2001} & - & 1.330 & - & 0.171 & -\\
Kit et al., 2004 \cite{kit2004} & - & 1.365 & - & 0.167 & -\\
Wu et al., 2014 \cite{wu2014} & - & 1.335 & - & 0.173 & -\\
\hline
mean & 0.715 & 1.343 & 0.236 &0.166 & 117.51 \\
$\sigma$ (std) & 0.012& 0.037 & 0.012 &  0.003 & 0.69\\
\hline
\hline
Present study & 0.698 & 1.367& 0.229 & 0.168 &116.95\\
\hline
\end{tabular}
\end{center}
\end{table} 

All quantities except $-\overline{c}_{pb}$ lie within the $1 \sigma$ interval of the mean reported value. The negative base pressure coefficient deviates slightly and is accurate to within $2 \sigma$. 

Additional details on CFD settings and results of a mesh convergence test can be consulted in \cite{decuyper2018} where an identical setup and configuration was used.
}

\section*{References}
\bibliographystyle{elsarticle-num} 
\bibliography{aerobib}

\end{document}